\documentclass[trackchanges,preprint2]{aastex7}

\usepackage[normalem]{ulem}
\usepackage{graphicx,xcolor}
\usepackage{subcaption}

\begin{document}

\newcommand{\vdag}{(v)^\dagger}
\newcommand{\fudesig}{{FUOr-Aur 0544+3330}}  %FO-Aur FOO-Aur FUO-Aur?
\newcommand{\obj}{ZTF 19abymxrr}  % aka WTP10aaktsc
\newcommand{\prog}{WISE J054452.25+333009.6} 

\newcommand{\lah}[1]{{\color{brown} #1}}
\newcommand{\lahcomm}[1]{{\color{brown} COMMENT: #1}}
\newcommand{\pla}[1]{{\color{blue}\ensuremath{***~} #1\ensuremath{~***}}}
\newcommand\done{{\color{purple}\ensuremath{\checkmark~}}}

%\newcommand{\deleted}[1]{\textcolor{red}{\sout{#1}}}
%\newcommand{\added}[1]{\textcolor{blue}{#1}}
%\newcommand{\replaced}[2]{\textcolor{red}{\sout{#1}}\textcolor{blue}{#2}}

%submitted to ApJ 
%VERSION:{\today}
%\received{28 February, 2025}
%\revised{, 2025}
%\accepted{\today}

%% Command to document which AAS Journal the manuscript was submitted to.
%\submitjournal{ApJL}

%\watermark{DRAFT}

%%%%%%%%%%%%%%%%%%%%%%%%%%%%%%%%%%%%%%%%%%%%%%%%%%%%%%%%%%%%%%%%%%%%%%%%%%%%%%%%
%\graphicspath{{./}{figures/}}

\title{FUOr-Aur 0544+3330: A New YSO Outburst \\ in the Outskirts of Auriga OB1, Viewed Face-On} 

\author{Lynne A. Hillenbrand} 
\affiliation{Department of Astronomy, MC 249-17, California Institute of Technology, Pasadena, CA 91125; USA}
\email{lah@astro.caltech.edu}

\author[0000-0002-9540-853X]{Adolfo S. Carvalho}
\affiliation{Department of Astronomy, MC 249-17, California Institute of Technology, Pasadena, CA 91125; USA}
\email{carvalho@astro.caltech.edu}

\author[0000-0003-2686-9241]{Dan Stern}  %optical spectrum / palomar
\affiliation{Jet Propulsion Laboratory, California Institute of Technology, 4800 Oak Grove Drive, Pasadena, CA 91109; USA}
\email{Daniel.K.Stern@jpl.nasa.gov}

%\affil[XXX]{\orgdiv{Jet Propulsion Laboratory}, \orgname{California Institute of Technology}, \orgaddress{\street{4800 Oak Grove Drive}, \city{Pasadena}, \state{CA} \postcode{91109}, \country{USA}}}

\author[0000-0002-8293-1428]{Michael Connelley} %second infrared spectrum / irtf
\affiliation{Institute for Astronomy, University of Hawaii at Manoa, 640 N. Aohoku Place, Hilo, HI 96720, USA}
\email{msconnelley@gmail.com}

\author[0000-0002-4128-7867]{Facundo P\'erez Paolino} %first infrared spectrum / apo
\affiliation{Department of Astronomy, MC 249-17, California Institute of Technology, Pasadena, CA 91125; USA}
\email{fperezpa@caltech.edu}

\author[0009-0009-6392-4810]{Ahaan Shetty}  %application of sedfitter 
\affiliation{Department of Astronomy, MC 249-17, California Institute of Technology, Pasadena, CA 91125; USA}
\email{ashetty@caltech.edu}

\author[0000-0001-5683-0095]{Zachariah Milby}  %donated time for HIRES/keck 
\affiliation{Department of Planetary Sciences, MC 150-21, California Institute of Technology, Pasadena, CA 91125, USA}
\email{zmilby@caltech.edu}

\author[0000-0002-0531-1073]{Howard Isaacson}
\affiliation{Astronomy Department, University of California, Berkeley, CA 94720, USA}
%\affiliation{University of Southern Queensland, Toowoomba, QLD 4350, Australia}
\email{hisaacson@berkeley.edu}

\begin{abstract}
We present a newly appreciated FU Ori outburst event 
that began in 2019 and reached a peak in early 2021.
Suspected young stellar object WISE J054452.25+333009.6 experienced substantial brightening,
in excess of $-5$ mag at optical wavelengths and $-2.5$ mag at mid-infrared wavelengths.
The time from near-quiescence to peak brightness was approximately 24 months.
Optical and near-infrared spectra confirm that the outbursting source
(hereby designated \fudesig) shows all the hallmarks of the FU Ori class,  
including the \ion{Li}{1} indicator of stellar youth. 
The mix of ionized and neutral atomic lines, 
alongside prominent molecular absorption features,
is consistent with the expected change in spectral type 
from earlier in the optical to later-type in the near-infrared.  
The closest analog among well-studied FU Ori objects is V1515 Cyg.  
Both sources have unusually narrow-lined absorption spectra that can be explained 
by a face-on disk orientation, such that disk-broadening is minimized and 
wind-induced blueshift (in e.g. H$\alpha$, NaD, \ion{Ca}{2}) is maximized.
Both the optical through infrared spectral energy distribution 
and high-resolution spectrum are well-fit by a pure-accretion disk model.  
Adopting a distance of $d=1.5$ kpc, the accretion and central star parameters are:  
$\dot{M} = 10^{-5.48}$ $M_\odot$ yr$^{-1}$, $M_* = 0.17 \ M_\odot$, and $R_\mathrm{inner} = 1.04 \ R_\odot$. 
Other fitted values are disk inclination $i=5.9$ deg 
and source extinction $A_V=1.83$ mag. 
These parameters yield accretion luminosity $L_\mathrm{acc} = 8.4\ L_\odot$ and 
maximum disk temperature $T_{\rm{max}} = 6218$ K.
%In order to rectify the post-outburst disk with the pre-outburst spectral energy distribution,
%we hypothesize that the outburst originated from the secondary star in a young Class I binary system.
\end{abstract}


\section{Introduction}
\label{sec:intro}

The astronomical community has benefitted over the past two decades from a plethora of 
wide-field photometric surveys conducted in the time domain. 
A main aim of many such surveys has been to catch rare events as they happen, in real time, 
and identify or localize ``transient" sources for multi-wavelength follow-up observations.
The literature is now replete with thousands of promptly identified supernovae, stellar mergers, 
tidal disruption events, and optical detection of transients from alert streams
such as gamma ray and gravitational wave events.  

Within the Galaxy, there has been industrial scale study of persistent 
stellar variables. These are mostly periodic sources with  regular lightcurve modulation 
caused by stellar rotation,  pulsation, or binary phenomena. 
%{\bf {REF??}}
But there are also transient brightening
phenomena, usually related to accretion, e.g. novae events, 
cataclysmic variables, and young stellar object outbursts, to name a few categories.
%{\bf {REF??}}

The variability in young stellar objects (YSOs), even when restricted to only
accretion-related origins, spans a wide range of amplitudes and timescales.  
At low amplitudes of $<1-10$ percent, and shorter timescales of hours to days, 
there is both stochastic variation and also discrete bursting, with duty cycles of days to weeks.  
NGC 2264 has been well-studied by e.g. \cite{cody2014,stauffer2014,stauffer2016} using CoRoT 
and Upper Sco e.g. by \cite{cody2017,cody2018} using Kepler/K2.
The observed behavior is likely associated with regular accretion processes, 
in which variable infall of material is mediated based on the strength and geometry of the magnetic field
\citep{zhu2025,zhu2024}.  
Intermediate burst amplitudes of a few mag, with durations of 
a few months, are called EX Lup-like bursts, while those lasting years
are called V1647 Ori-like events. 
For these, the duty cycles are years to decades, and they are likely caused by instabilities 
associated with the magnetospheric region, e.g. pinching and reconnection.
Finally, high amplitude and long-duration outbursts are termed 
FU Ori eruptions.  These are more catastrophic disk instability driven events 
for which the true duty cycles and event rates are not well-quantified.
See \cite{herbig1977,hartmann1996} and \cite{fischer2023} for reviews of
these various categories of YSO outbursters. 

Members of the FU Ori class, the most extreme YSO accretors in the post-initial infall stage,
have several distinguishing characteristics.
Their optical spectral types are FGK-type, while near-infrared spectra 
are M-type, with low surface gravity signatures throughout.
The wavelength-dependent temperature pattern includes absorption line profiles 
that are generally broad and non-Gaussian, consistent with those of a rotating disk.
FU Ori type stars generally exhibit strong wind/outflow signatures in lines such as
H$\alpha$ and sometimes higher Balmer lines, as well as the \ion{Ca}{2} triplet, and \ion{He}{1} 10830 \AA,
which all tend to show P Cygni or blue-shifted absorption.

In this paper, we present evidence that the unstudied 
YSO candidate \prog\ underwent an accretion-driven eruption, 
A multi-wavelength lightcurve spanning several years 
samples the pre-outburst, outburst, and outburst plateau phases.  
Recent spectroscopy at both low and high spectral resolution 
provide confirming evidence that the source exhibits all the usual 
spectroscopic requirements for FU Ori status, including a multi-temperature spectrum 
and strong wind signatures.  In addition, the 0.36-5.0 $\mu$m spectrophotometric and spectral properties 
are very well-modelled by an accretion disk model.
We conclude that the source is a newly discovered example of the FU Ori class of YSOs.  

We also propose a novel naming system for newly discovered or newly confirmed FU Ori objects,
with the new FU Ori object that we present here designated \fudesig. 

\section{The Pre-Outburst Source}
\label{sec:pre}

\subsection{Region and Distance}

The source of interest is located at RA = 05:44:52.25, Dec. = +33:30:09.6 (J2000), in the constellation of Auriga. It
is within the nominal boundaries of the Aur OB1 association, but adjacent to regions thought to be part of the more distant Aur OB2 association.  These groupings of HII regions and young massive open clusters are likely associated with the local spiral arm (Aur OB1) and the Perseus spiral arm (Aur OB2)
based on the discussion below regarding distances, and considering the material in \cite{hou2021}.

The immediate vicinity of the source is not particularly notable in terms of nebulosity or clustering.  No other YSO candidates are identified within 15 arcmin \citep{wenger2000}.
The object is a few degrees south of the string of HII regions Sh 2-231 to 235, and a few degrees east of major star forming region NGC 1931 (Sh-2 237), as well as north-northeast of AFGL 5157.  %05:37:48 +31:59:24 
\cite{reipurth-yan2008} provide a more complete description of the overall region, and the star formation activity it harbors.  

In terms of previous literature, the source appears only in a catalog paper of the Auriga region by \cite{pandey2020}, where it is source \# 130. However, it appears to have been beyond the edge of their new imaging survey area and is not in the published catalog\footnote{Detection limits were reported as $V<21.2$ and $I<20.0$ so the source should have been present in the optical catalog if covered by the imaging survey.}.  It is therefore exterior to both of the large ISM bubbles that they identify.  Nevertheless, \cite{pandey2020} characterize the pre-outburst source (\prog) as a Class I type YSO based on its mid-infrared colors alone.  The nearest similarly identified YSOs are located approximately 15 arcmin away, in projection.

Regarding distance, \cite{pandey2020} assumed 2.2 kpc for their region under study.  However, at least some of the star formation activity in the general area is at closer distance.  Specifically, some of the Sharpless regions as well as LDN 1525 a few degrees to the north are at around 1.3 kpc \citep{straizys2010}.  Also nearby in projection (to the northwest) is the well-studied open cluster NGC 1960 ($\sim$15 Myr) located $1.2\pm0.13$ kpc away \citep{panja2021} while AFGL 5157 a few degrees to the south-southwest is at 1.6 kpc \citep{verma2024}.   Further detailed study of the 3D distribution of the stellar clustering in this broad area of the sky is provided by \cite{quintana2023}.  Consistent with early analysis by \cite{humphreys1978}, Aur OB1 and Aur OB2 are located at different distances.  
\cite{melnik-dambis2020} find distances of 1.06 and 2.42 kpc, respectively, 
though as pointed out by \cite{quintana2023}, neither Aur OB1 nor Aur OB2 is a particularly coherent kinematic group. Each appears to have about -50\% to +100\% range in distance, i.e. the associations may be partly overlapping along the line of sight.

Gaia DR3 does not provide a reliable distance for the source itself, measuring a negative parallax ($\pi = -0.18 \pm 0.19$).  This loosely suggests a minimum distance of about 1.7 kpc, and perhaps consistency with the estimate for the further Aur OB2.
Gaia also reports proper motion measurements  $\mu_{R.A.}cos(Dec.) = -2.37 \pm 0.22$ and $\mu_{Dec.} = 0.17 \pm 0.12$ mas/yr.  Although not definitive, this motion seems more consistent with Aur OB1 than Aur OB2 members \citep{quintana2023}.  Furthermore, the nearest-in-projection stars studied by  \cite{quintana2023} are members of their sub-groups 1 and 3, with median distances 1.1-1.5 kpc.

Based on sky position, proper motions, and luminosity consistency with pre-main sequence stars, 
we might suspect that the ``near" Aur OB1 distance is the appropriate location for the outburst source. 
%As discussed below, we find from disk model fitting of the outburst that $d=1.65$ kpc is the most likely distance.
In what follows, we adopt a distance of 1.5 kpc. Supporting evidence for the plausibility of a YSO population 
at such distance comes from the 3-D dust extinction maps provided by \cite{edenhofer2024}, 
with $14\arcmin$ spatial resolution, which show a significant dust enhancement in this distance range.

\begin{figure}[!t]
\includegraphics[width=0.85\linewidth,angle=-90]{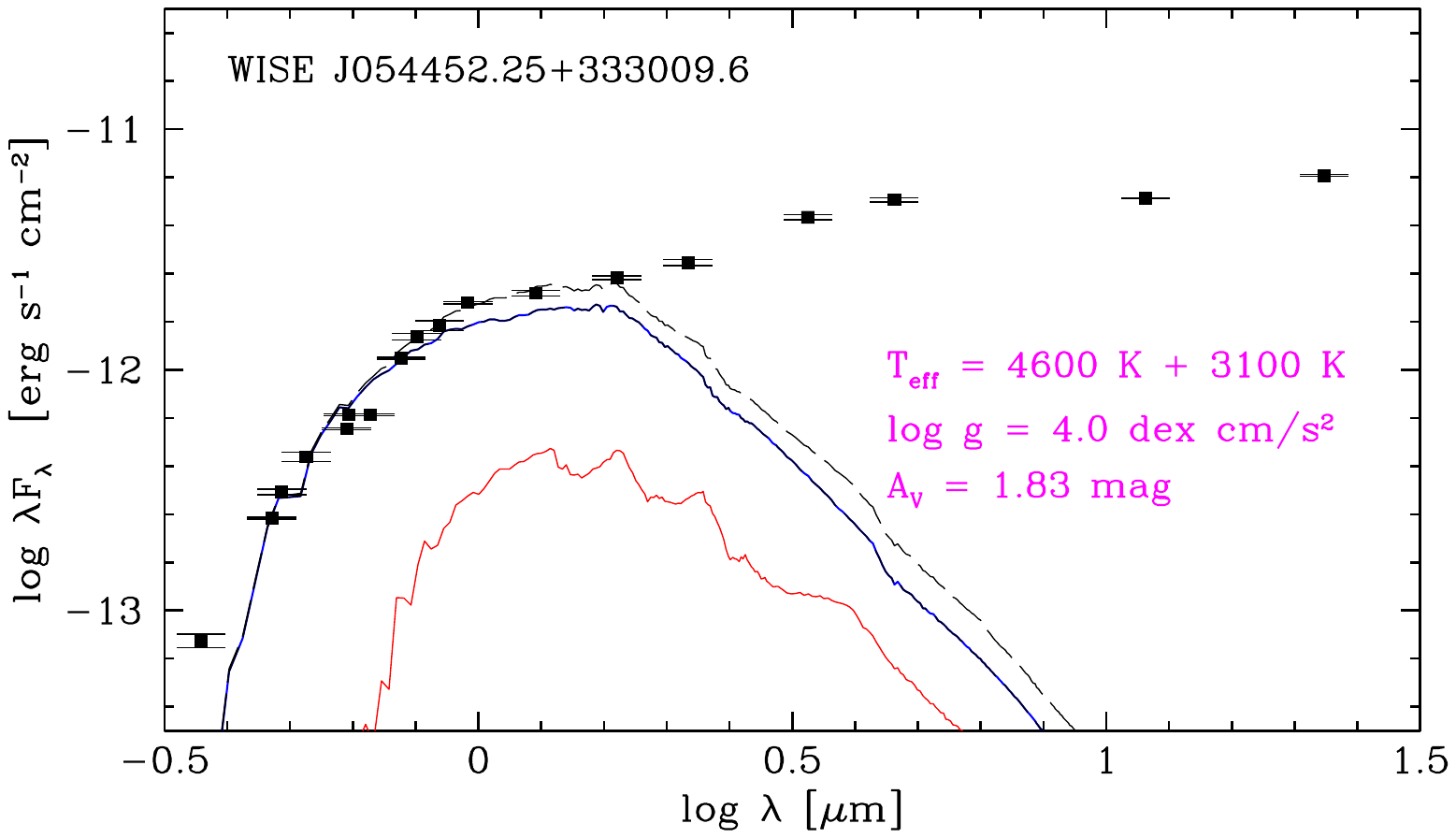}
\vskip-0.5truein
\caption{
Spectral energy distribution showing the pre-outburst photometry (black points) 
%and outburst (blue points) 
along with a two-star model atmosphere \citep{allard}. Both the model stars 
have log g = 4.0 dex cm/s$^{2}$ and are extincted by $A_V=1.83$ mag. 
One has $T_{eff}=4600$ K (blue line) while the other has $T_{eff}=3100$ K (red line) and 10\% the flux of the first 
at the normalization wavelength in the red optical.  The dashed line is the sum of the two models.  
The binary source exhibits both an ultraviolet/blue excess and an infrared excess
relative to this binary model. The excesses can not be eliminated through a 
different combination of stellar and extinction parameters.  The SED
is a Class I type, with $\alpha=+0.36$, and consistent with a YSO
surrounded by an accreting and likely flared circumstellar disk, possibly with a thin envelope.
}
\label{fig:obssed}
\end{figure}
\subsection{Progenitor Properties}

A pre-outburst spectral energy distribution (SED) of \prog\ can be constructed using photometry from: 
USNO-B \citep{monet2003},
Gaia \citep[$G,B_P,R_P$; ][]{gaiadr3}, 
UVEX \citep[$U,g,r$; ][]{monguio2020}, 
iPHaS \citep[$r,i$; ][]{barentsen2014}, 
PanSTARRS \citep[$g,r,i,z,y$;][]{flewelling2020}, 
2MASS \citep[$J,H,K$;][]{skrutskie2006}, 
and WISE \citep[$W1,W2,W3,W4$;][]{wise_data}. 
We note that the source does not appear in the 2MASS PSC (Point Source Catalog), only the XSC (Extended Source Catalog).
We therefore compute original $J,H,K$ magnitudes from 2MASS Atlas images 
downloaded from the IPAC/IRSA Image Service. This photometry is described in Appendix \ref{app:2MASS}.

The resulting SED is shown in Figure~\ref{fig:obssed}. 
Its shape in the optical wavelength range requires a star warmer than about 4000 K.
However, the composite stellar model shown in Figure~\ref{fig:obssed} was 
guided by our accretion disk model of the outburst source, which is described in a later section.
In the disk model fitting, the constrained parameters are the stellar mass $M_*$ (which implies a temperature $T_{eff}$) 
and the stellar radius $R_*$.  Together, they produce a luminosity that is a factor of several too low compared to the observed pre-outburst SED.
To resolve the luminosity discrepancy, we hypothesize that the outbursting source is a binary and the secondary star is the one that has undergone the outburst.

Figure~\ref{fig:obssed} illustrates that the progenitor SED
can be approximately matched by a composite stellar atmosphere model \citep{allard}
having a $T_{eff}=4600$ K primary and a $T_{eff}=3100$ K secondary 
with 10\% of the primary's flux at the normalization wavelength.
We adopt log g = 4.0 dex g cm$^2$/s as representative and $A_V\approx 1.83$ mag,
which results from our later disk model fitting but is consistent with independent estimates 
of 1.5-2 mag from the SED.
The luminosity implied from such stellar atmosphere models is 
$5.0\times (d/1.5 \mathrm{kpc})^2 L_\odot$ for the primary and $0.55\times (d/1.5 \mathrm{kpc})^2 L_\odot$ for the secondary. 
Such a secondary progenitor is consistent with the $M_*$ and $R_*$ from the outbursting accretion disk model.  
%{\bf [yay adolfo for figuring this out!]}.
And the luminosity of the combined stellar model -- that includes both the outburst progenitor star
and a hotter companion star -- provides a reasonable fit to the optical and near-infrared portion 
of the pre-outburst SED, especially when allowing for contribution from gas/dust in an inner disk.
A model comprised only of the secondary source is unable to match the SED. 

The progenitor (binary) source shows both large infrared excess and some blue optical/ultraviolet excess.   
The blue excess can be attributed to accretion, while the red excess is interpreted as circumstellar dust.
In order to fit the near- and mid-infrared part of the pre-outburst SED, 
warm dust over a range of temperatures is required.
%warm dust at $\sim800$ K plus cooler dust ranging down to $220$ K is required.  
There are no available measurements at wavelengths longer than WISE $22.4 \mu$m.  The source
\prog\ has a Class I type SED with spectral index +0.36 (rising to the red).
This is consistent with a pre-main sequence star surrounded by a highly flared disk 
or a circumstellar envelope, which is the geometry needed 
in order to reproduce the relatively flat mid-infrared spectral slope \citep{robitaille2017}.
Appendix ~\ref{app:sedfitter} presents one version of such a model.

In addition to the SED with its infrared and blue/ultraviolet excess, 
there is evidence of H$\alpha$ emission in \prog.  This is
based on consideration of iPHaS photometry and specifically the H$\alpha - r$ vs $r-i$ color 
%  r-i = 18.33-17.29=1.02 and r-Ha = 18.33-17.27 = 1.05 
indices \citep{barentsen2014}. 

Overall, we conclude that the progenitor source \prog\ exhibited the characteristics 
of a Class I type YSO.
The immediate environment in which it is located, however, is not an obviously active star forming region. 

\section{Outburst Source Lightcurve}

\begin{figure}[!t]
\includegraphics[width=\linewidth]{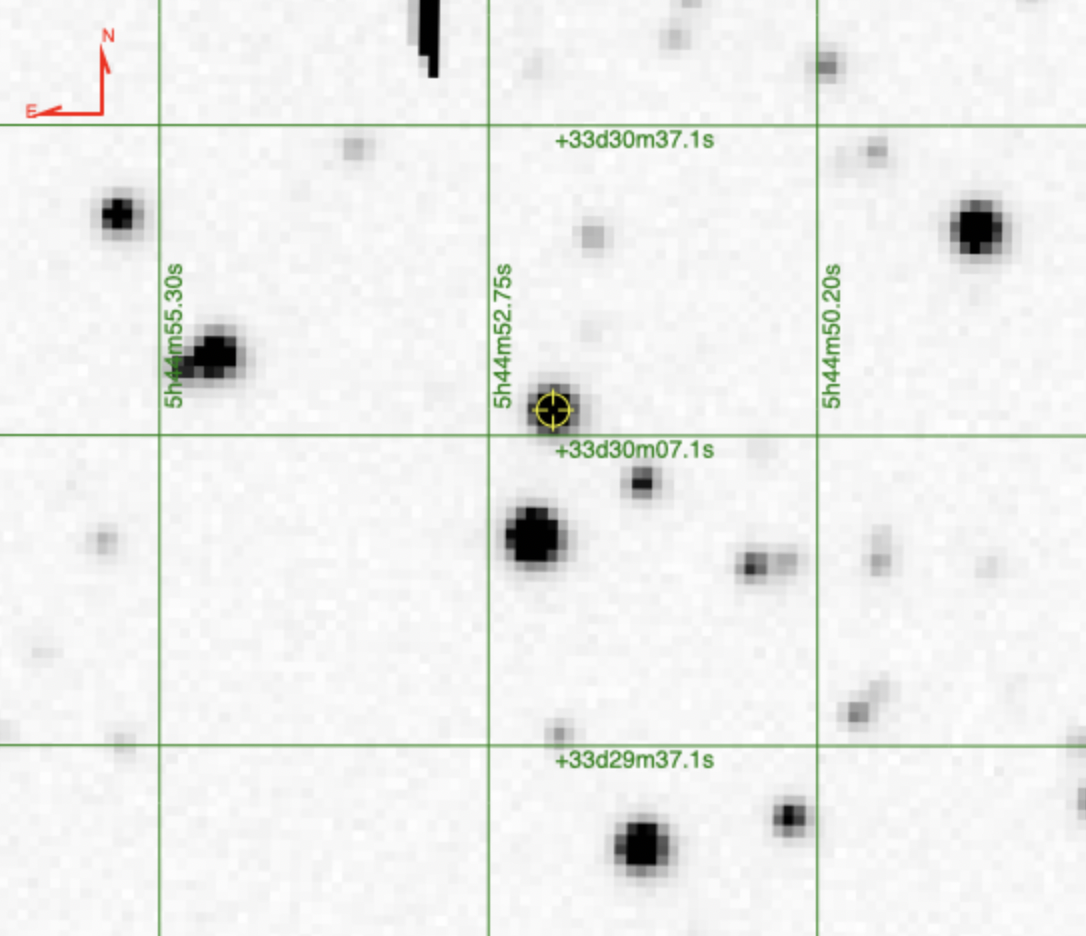}
\caption{
Image retrieved from the IPAC ZTF Image Access service\footnote{\url{https://irsa.ipac.caltech.edu/Missions/ztf.html}} showing \obj\ (bullseye, near center) in its current outburst state.
}
\label{fig:image}
\end{figure}

\begin{figure}[!t]
\includegraphics[width=1.15\linewidth]{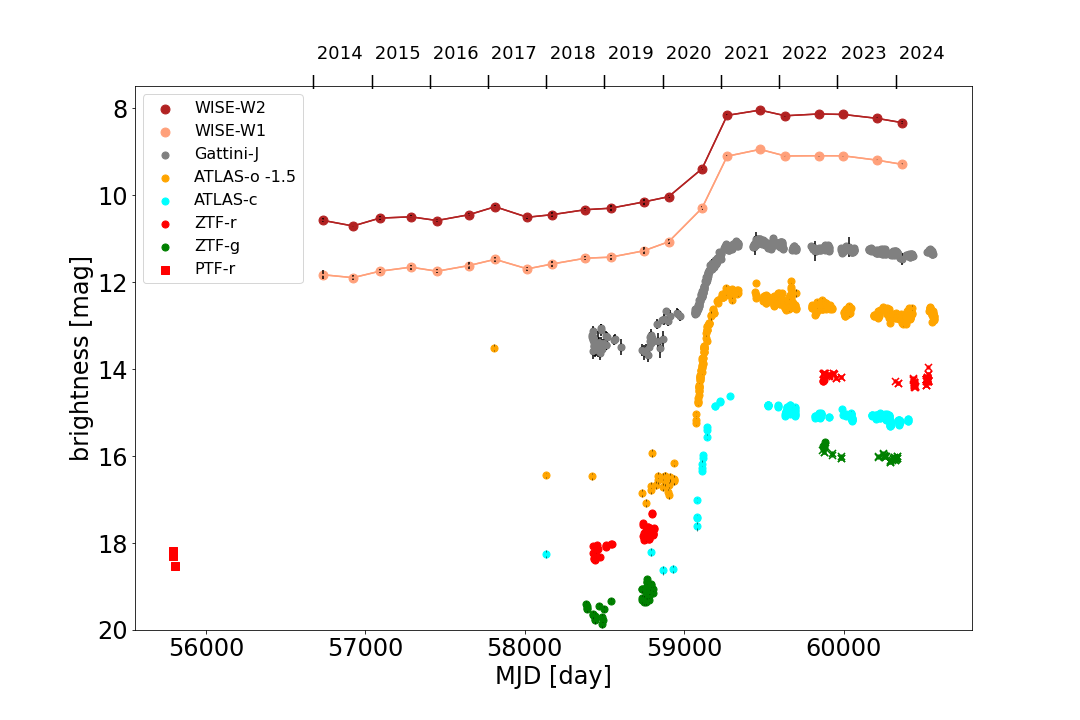}
\includegraphics[width=1.15\linewidth]{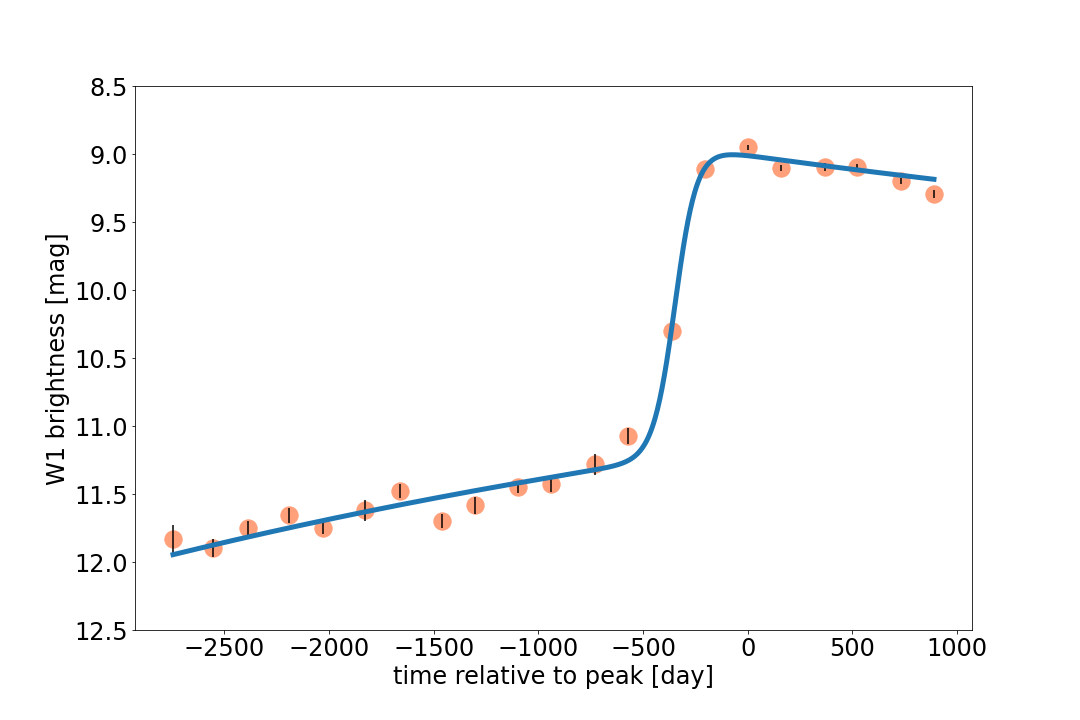}
\caption{
\textbf{Top:}
Multi-wavelength lightcurves of \obj, sampling the pre-outburst low-accretion state,
the outburst, and the current high plateau phase.
The lightcurve appears to have peaked in mid-2021 and has
been slowly and steadily declining since that time. \textbf{Bottom:} Functional fit to the NEOWISE $W1$ lightcurve.
}
\label{fig:lc}
\end{figure}

Lightcurve data for the outburst source was assembled from:
ZTF \citep[Zwicky Transient Facility; ][]{bellm2019} and its predecessor,
Palomar Transient Factory \citep{law2009}, as accessed through IRSA \citep{masci2019},
ATLAS \citep[Asteroid Terrestial-impact Last Alert System; ][]{tonry2018},
Gattini \citep{de2020,gattini_data},
and NEOWISE \citep[Near Earth Object WISE reactivation;][]{mainzer2014}, as accessed
through IRSA\footnote{{\url{https://wise2.ipac.caltech.edu/docs/release/neowise/}}} \citep{neowise_data}.
A ZTF image of the outburst source is shown in Figure~\ref{fig:image} and
the multi-wavelength lightcurve in Figure~\ref{fig:lc}.

The source has historical brightness of $r\approx 18.2-18.5$ mag in Palomar Transient Factory measurements from 2011,
consistent with pre-outburst measurements in late 2018 of $r\approx 18$ and $g\approx 19.5$ mag 
in Zwicky Transient Facility time series data. In the 2018-2019 season, the source began brightening, 
which triggered the ZTF designation \citep{graham2019} \obj. 
The object was observed by ZTF again in 2024, having transitioned to a bright state of $g\approx 15.5$ and  $r\approx 14.2$ mag.   
A sigmoid fit to the r-band lightcurve characterizes a rise amplitude of -5.08 mag.    The rise time is not well-constrained due to lack of sampling of the actual rise by ZTF.

The outburst is also detected in the Palomar Gattini infrared time domain survey data. A sigmoid fit to the $J$-band lightcurve characterizes a rise amplitude of -2.30 mag.  However, this is a lower limit given the Gattini pixel size (8.7$\arcsec$) and the certainty that the aperture photometry includes a bright star located $\approx 10\arcsec$ to the northeast that thus biases both the faint-state and the bright state measurements of \fudesig.   From consideration of the faint state 2MASS measurement (\S\ref{app:2MASS}) 
relative to the outburst spectrophotometric measurements (described below), the true $J$-band amplitude is considerably larger, $\Delta J = -3.9$ mag. 
The fitted rise time is 443 days (14.8 months), although this neglects the shoulder of -0.66 mag before the final rise, so the true rise timescale is somewhat longer, by about 261 days (8.7 months).  The total rise time is therefore 704 days (23.5 months).

In data from WISE and NEOWISE, the source brightening can be characterized over a 14 year timescale.   A shallow rise phase can be mapped beginning in 2014 and through early 2020, characterized by $-0.083$ mag/year brightening that then steepened in late 2020 through early 2021 to $-2.80$ mag/year (-0.23 mag/month). 
A sigmoid function fit to the W1 lightcurve reveals a -2.69 mag rise over 738 days (24.6 months) while the same for W2 results in a -2.48 mag rise over 707 days (23.6 months). This fit, with $\chi^2=12$, does not capture the behavior of the early NEOWISE lightcurve. A better fit is provided by a function that incorporates both the shallow rise and the subsequent steeper rise -- a morphology that is common to several recently measured FU Ori outbursts, though not ubiquitous.    
The functional form
$$a~\mathrm{tanh}((t - b) / c) \times e^{-(t - b) / d} + \rm{offset} $$ has $\chi^2=3.95$ for parameters:
$a= -1.15; \ 
b= -345.24; \ 
c= 103.70; \ 
d= 4958.38; \ 
\mathrm{offset} = 10.08$, where $t$ is the time in days relative to lightcurve peak. This fit is shown in Figure~\ref{fig:lc}.

The peak brightness in this model fit to the mid-infrared NEOWISE data
occurred around MJD$\approx 59400$ days, in early July of 2021. 
The near-infrared peak appears to occur at a similar time.
The optical peak, however, may occur somewhat earlier, perhaps by a few months
(see Figure~\ref{fig:lc}). 

We adopt approximately 1.3 years as the main rise time, based on our functional fit.
This timescale for the main brightening is roughly consistent between the photometric bands  
from optical to mid-infrared. We emphasize the precursor, shallow rise, 
that spanned at least five years before the main brightening in 2020-2021.
 
We can also consider the color-magnitude evolution of the source.  The optical color-magnitude evolution is fairly neutral in $g-r$, 
with the outburst perhaps even slightly redder than the historical $g-r\approx 1.4$ for the source.
In the infrared, however, during the lightcurve rise the $W1-W2$ color becomes bluer, by about 0.25 mag from quiescence to outburst peak. 
The initial shallow rise in late 2018 and early 2019 is accompanied by a gradual blueward drift by about 0.1 mag. 
During the sharper rise beginning in 2020 and into 2021, the color jumps by 0.15-0.2 mag blueward.

\section{Outburst Source Spectroscopy}

Above we have referred to the subject of this paper by the name of the
quiescent source, \prog, and then by the designation given to it
when a photometric alert was issued, \obj.  Subsequently when discussing the spectroscopy,
which confirms the object as a bona fide FU Ori outburst, we shall use the name \fudesig\
as mentioned in the Introduction (\S~\ref{sec:intro}) and detailed in the Discussion (\S~\ref{sec:disc}).

\subsection{Data Acquisition}

\subsubsection{Optical}

A low-resolution optical spectrum was obtained by D. Stern on 2024-09-11 (UT) with the Palomar Double Spectrograph \citep{oke1982,kirby2011} and spans approximately the full optical range. Integration times of 900 s (blue) and $3\times 250$ s (red) were used.
%and produced excellent signal to noise.   % {\bf quantify "excellent signal-to-noise"}
After data reduction and spectrum extraction, flux calibration was performed.  

Two high-resolution optical spectra were obtained, both with the W.M. Keck Observatory.
First, the Keck I telescope and HIRES spectrometer \citep{vogt1994} were used on 2024-09-17 (UT)
by L. Hillenbrand and Z. Milby.  %images 142, 143, 144 with flats/bias in afternoon; OA Rita
Three exposures (750 s, 400 s, 400 s) were obtained with the C5 decker (1.15"). 
Data reduction was performed using the PypeIt spectral extraction package \citep{prochaska2020},
resulting in a S/N =30-50 spectrum\footnote{These estimates are from dividing the spectrum by a 20 pixel median-filtered version of itself and computing the standard deviation in the middle 50\% of the order.}
at resolution $\approx 36,000$  over $\approx 5200-9700$ \AA\

Second, on 2024-10-03 (UT) H. Isaacson followed up with Keck Planet Finder \citep[KPF; ][]{gibson2024} spectroscopy, obtaining three exposures of 1800 s.  
Data reduction was performed by the automatic pipeline through the CPS (California Planet Search) 
collaboration. The final data product contains three apparitions of the source and one sky background spectrum. Each source spectrum has a slightly different flux level due to the differently sized slices of the KPF science fiber. To account for this and to properly subtract the sky background, we scale the sky spectrum by the median value of each science spectrum, then subtract. 
Finally, the nine  background-subtracted apparitions of the source spectrum 
were each normalized to a median of 1 and combined by taking the noise-weighted mean.   
The final KPF spectrum has resolution $\approx 95,000$ and covers $\approx 4450-8700$ \AA, with $S/N=50$ at 5000 \AA\ and $S/N>100$ at $>6000$ \AA\ measured as above, near the center of the orders.

\subsubsection{Infrared}

A near-infrared spectrum was obtained on 2024-10-28 (UT) using the Apache Point Observatory's 
TripleSpec instrument \citep{wilson2004} 
and a 1\arcsec.1 slit,
by F. Perez Paolino, L. Hillenbrand, and B. Horner.   
Sixteen image quads in ABBA format were obtained using 180 s integration times.
The resulting spectra cover 0.96-2.45 $\mu$m at $R\approx 3500$.
%slit viewer imaging was also obtained - diff.fits -}

NASA's IRTF and SpeX instrument \citep{rayner2003} were used by M. Connelley on 2024-10-30 UT 
to obtain spectra in the LXD mode, covering the
$L$ and $M$ atmospheric windows (with good quality data from $\sim 3-5~\mu$m).
Dithered exposures were used to obtain a total on-source integration time of 8 min.

Then, on 2024-11-08 UT, SpeX was also used in its SXD mode by M. Connelley to obtain spectra 
in the $YJHK$ region ($0.7-2.5~\mu$m).
Dithered exposures were used to obtain a total on-source integration time of 16 min.
The 0.5\arcsec\ slit was used, yielding an effective spectral resolution of $R=1200$. 
$K_\mathrm{MKO}$-band images were also acquired using guider images, with aperture photometry resulting in a
photometric measurement $K_\mathrm{MKO}=10.31\pm 0.04$ mag.  This is brighter by $\Delta K = -3.5$ mag 
than the quiescent-phase 2MASS measurement (See \S\ref{app:2MASS})

For the APO/TSpec and the IRTF/SpeX (SXD and LXD) observations,
spectral images were reduced using the $Spextool$ package \citep{cushing2004}. 
Telluric correction and flux calibration were performed using the 
$XTellCor$ package \citep{vacca2003} and A0V-type spectral standards observed close in airmass
that are assumed to have the same throughput as the objects.

%{\bf APPARENTLY THERE IS ALSO AN MMT SPECTRUM TAKEN BY KDE, BUT I DO NOT HAVE IT}

\subsection{Description of Absorption Spectrum}

\begin{figure*}[!htb]
\includegraphics[width=0.8\linewidth, angle=-90, trim={0 0 0 0.85cm},clip]{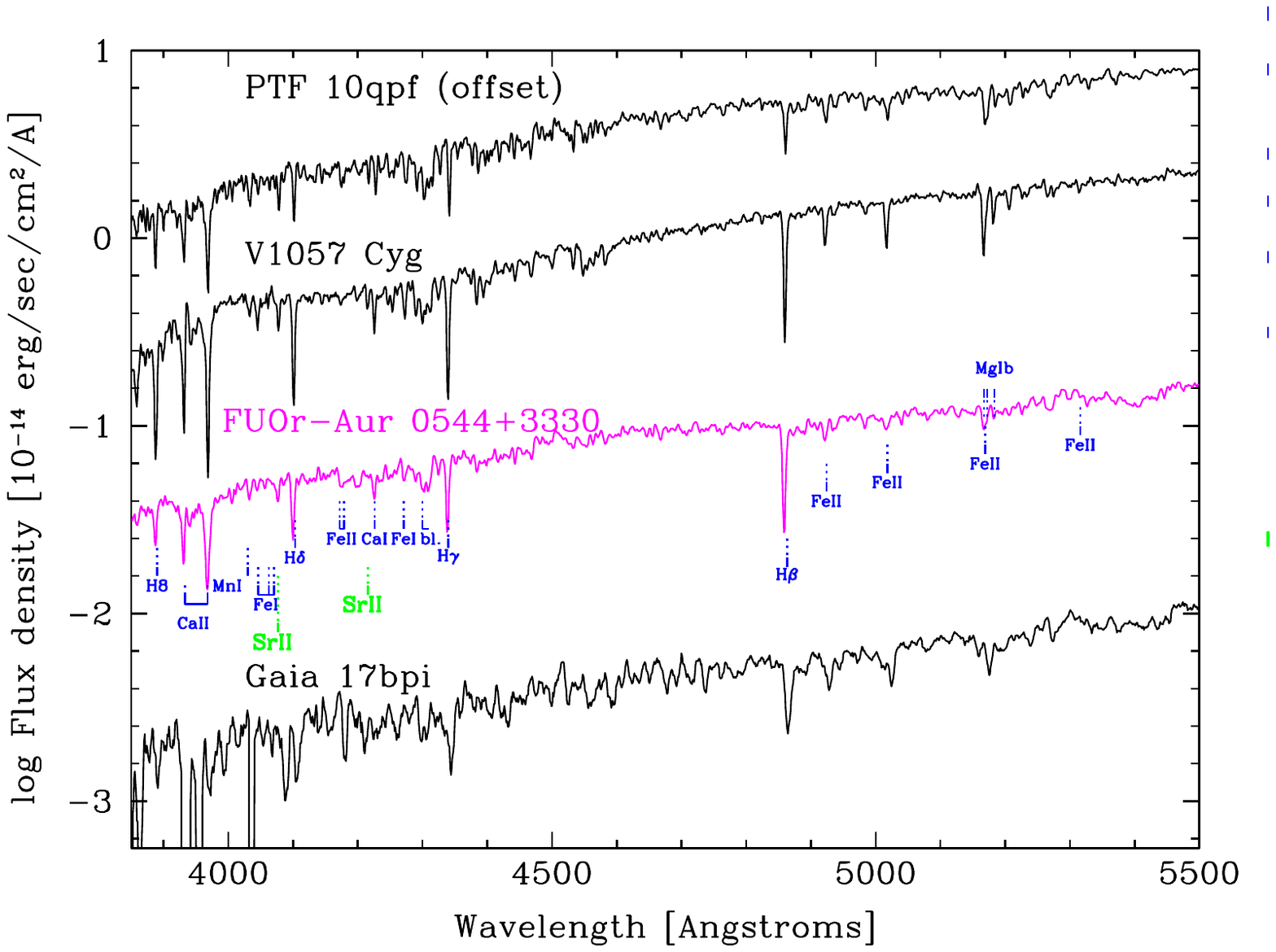}
\caption{
Blue portion of the lower resolution optical Palomar/DBSP spectrum of \fudesig\ (magenta) 
compared to the FU Ori objects 
PTF 10qpf (HBC 722), V1057 Cyg, and Gaia 17bpi.  
At low resolution, \fudesig\ has an absorption spectrum similar to the comparison objects,
both here in the blue as well as further to the red.
Deep Balmer lines and \ion{Ca}{2} lines are seen in addition to \ion{Mg}{1}b and \ion{Na}{1} D,
all formed in a wind.  Also present are a mix of \ion{Fe}{2} and \ion{Fe}{1}
as well as other metal lines that are formed in an accretion disk photosphere.
\fudesig\ thus appears to be an extreme accretor.
}
\label{fig:optspec}
\end{figure*}

\begin{figure*}[!t]
\includegraphics[width=0.5\linewidth,trim={0.85cm 0 0.85cm 0},clip]{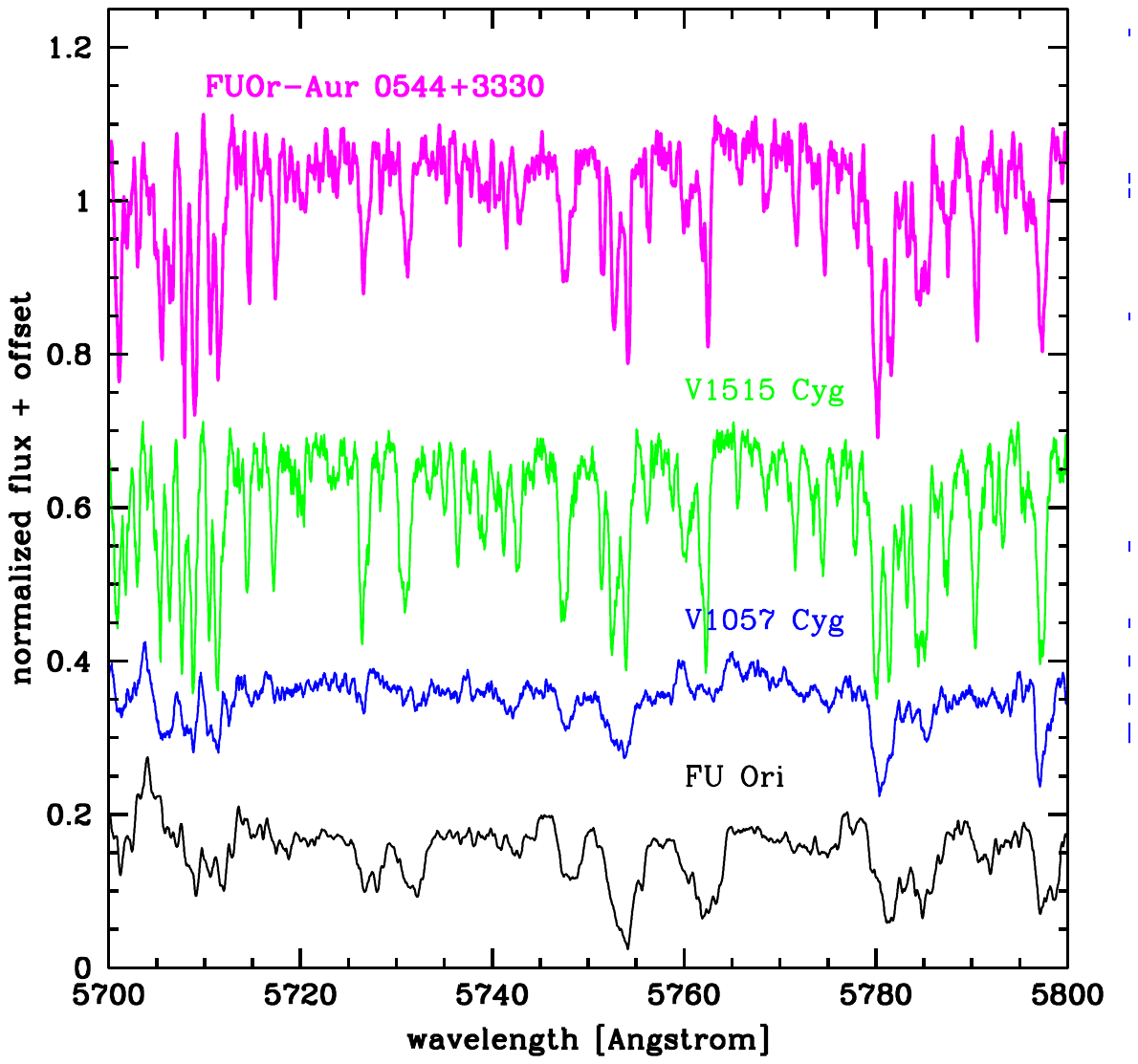}
\includegraphics[width=0.5\linewidth,trim={0.85cm 0 0.85cm 0},clip]{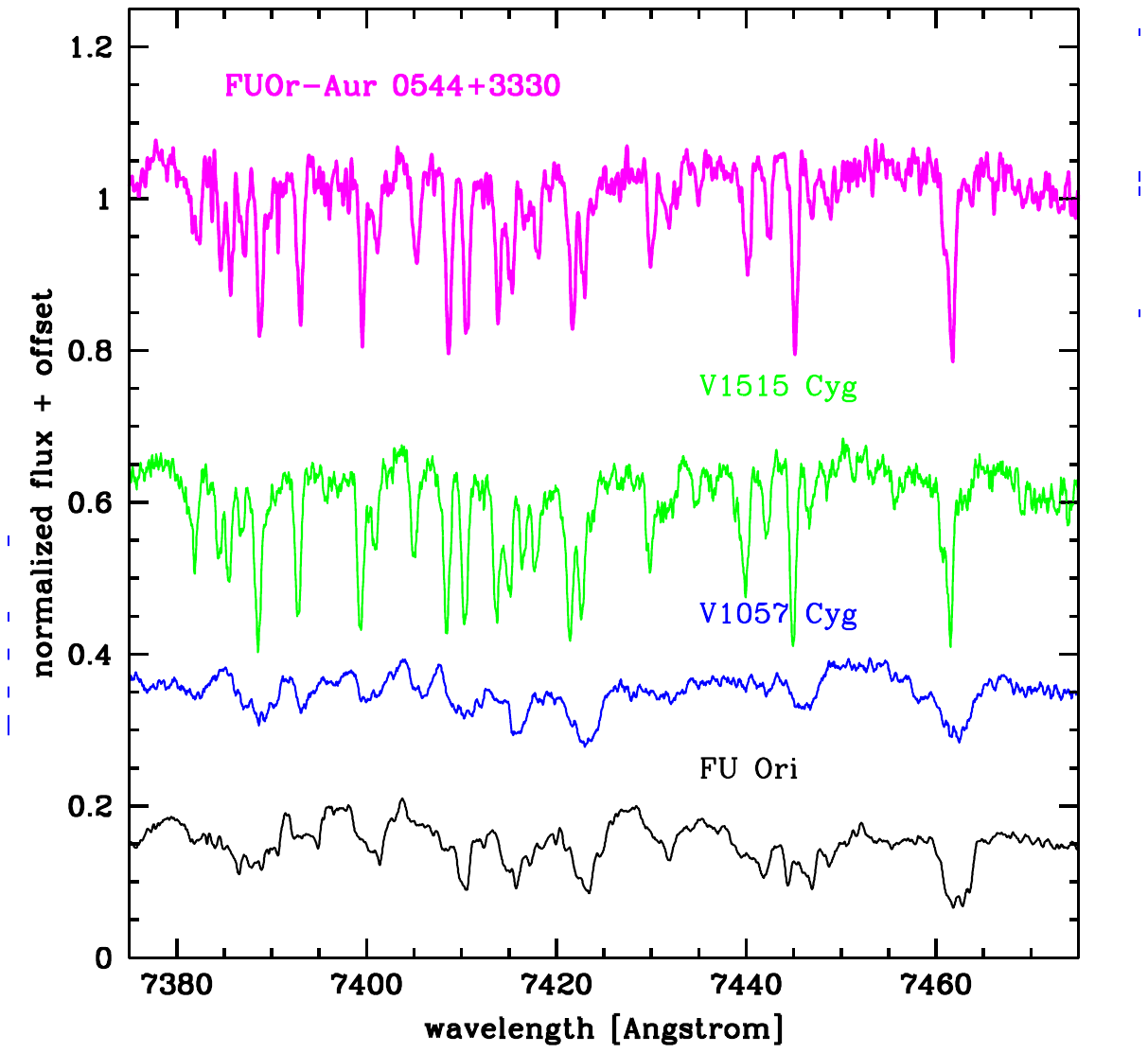}
\vskip-1.0truein
\caption{Two spectral segments from the higher-resolution optical Keck/HIRES spectrum 
of \fudesig\ (magenta), compared to the classic FU Ori objects 
FU Ori itself, V1057 Cyg, and V1515 Cyg.   The absorption line widths
are strikingly similar between \fudesig\ and V1515 Cyg, and both objects are much less 
(disk-)broadened than the lines exhibited by V1057 Cyg and FU Ori.}
\label{fig:compare}
\end{figure*}

%\begin{figure}[!t]
%\includegraphics[width=0.5\linewidth,trim={0 0 0 13cm},clip]{lines_ph.pdf}
%\vskip-1truein
%\caption{Lines featured by \cite{petrov2008} having a range of intermediate
%excitation potentials, as seen in our Keck/HIRES optical spectrum of \fudesig.
%} 
%\label{fig:ph}
%\end{figure}

\subsubsection{Optical}

\begin{figure}[!htb]
\includegraphics[width=1.05\linewidth,trim={0.85cm 0 0 0},clip]{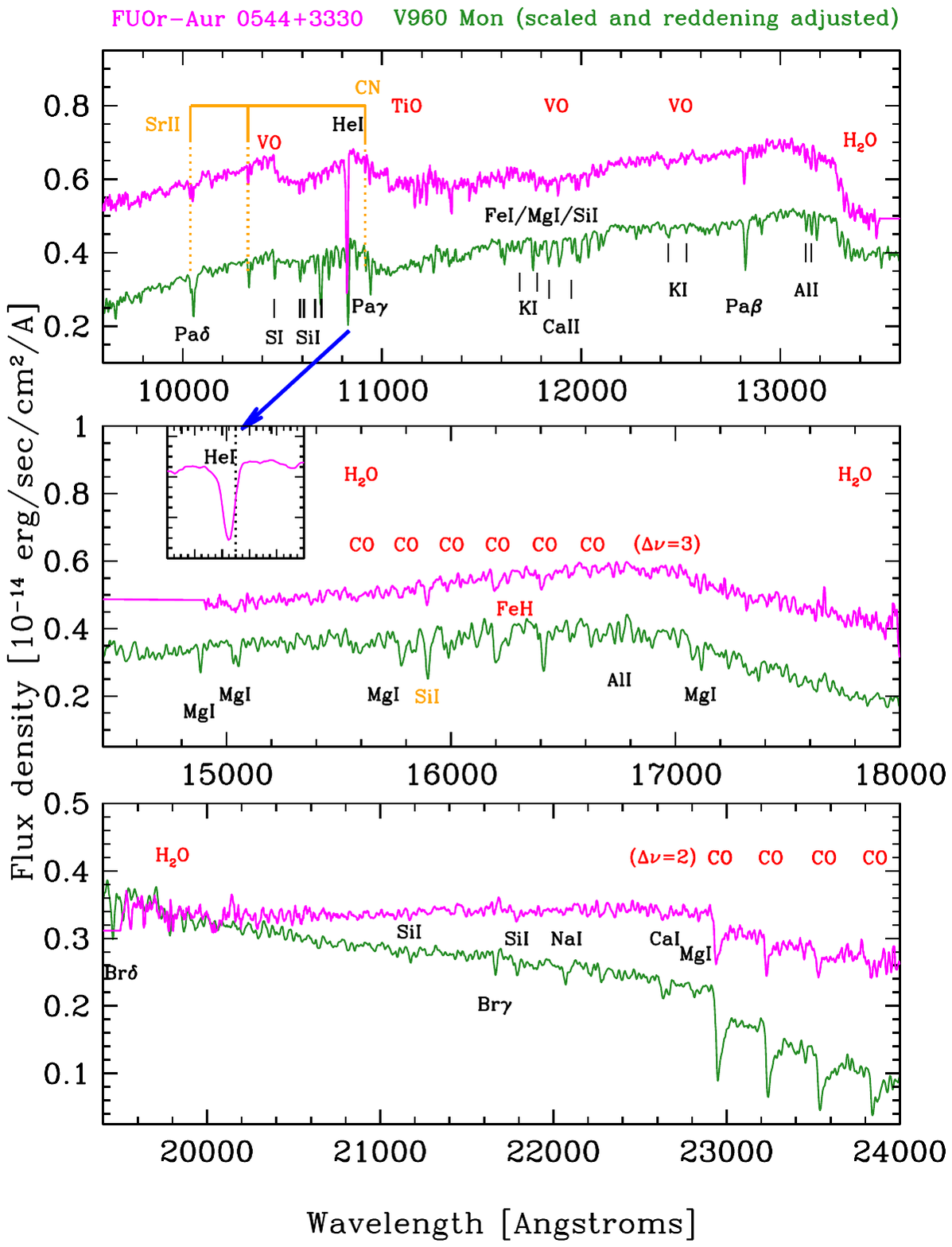}
\caption{ 
Infrared spectrum from APO/TSpec 
(higher spectral resolution albeit lower signal-to-noise relative to the IRTF/SpeX spectrum)
of outburst source \fudesig\ (magenta)
compared to V960 Mon (green), where the latter has been
artificially reddened to roughly match the continuum shape of the outburst source.
Both sources show similar absorption features from a mix of molecules (red)
%(TiO, VO, H$_2$O, CO; red), low-excitation (\ion{Mg}{1}, \ion{Ca}{1},  \ion{Al}{1}), 
%and higher-excitation (\ion{Si}{1}, \ion{S}{1}, \ion{C}{1}, \ion{Ca}{2}) neutral and ionized species (black), 
and lower- as well as higher-excitation atomic species (black).
Surface-gravity sensitive \ion{Sr}{2}, CN band, and \ion{Si}{1} features (orange) are also present.  
The line widths in \fudesig\ are notably narrower than those in V960 Mon.
We also call attention to the strong blueshifted absorption in \ion{He}{1} 10830 \AA\ (inset),
indicative of a strong wind.
}
\label{fig:irspec}
\end{figure}

The blue range of our low-resolution optical spectrum is shown in Figure~\ref{fig:optspec},
and several segments of the high-resolution optical spectrum appear in
Figure~\ref{fig:compare}, compared in both cases to other FU Ori type objects.
%Figure~\ref{fig:ph} highlights several individual atomic spectral lines having moderate excitation potential.
The full infrared spectrum is shown in Figure~\ref{fig:irspec}, 
also in comparison to an FU Ori type object.

At low dispersion, the optical spectrum of \fudesig\ exhibits the features expected from an FU Ori object.   Overall, it has a G-type appearance but with much deeper absorption lines in the Balmer series, \ion{Na}{1} D, as well as various \ion{Fe}{2} lines located in the blue.  Higher excitation species like \ion{O}{1} 8446 \AA\ and \ion{Mg}{1} 8807 \AA\ in the red are also present.  
In the high dispersion data, one sees intermediate excitation lines from 
e.g. \ion{Ni}{1}, \ion{Fe}{1}, and \ion{Fe}{2} (notably 5316, 6516 \AA), 
as well as lower-excitation species like \ion{Ti}{1} and \ion{Fe}{1}.  
The key \ion{Ba}{2} 6141 \AA\ line and 6496 \AA\ line complex are evident.   

Examination of H$\alpha$ and H$\beta$ reveals clearly blueshifted absorption profiles, even at low spectral resolution.   Similar blueshifted absorption is detected in \ion{Mg}{1}b, \ion{Na}{1} D, \ion{K}{1} 7699 \AA, and the \ion{O}{1} 7773 \AA\ triplet.  These lines all provide evidence of a strong outflow, as discussed in more detail below.
At high dispersion, we see strong absorption from high-excitation, hotter lines such as \ion{Fe}{2}, 
e.g. \ion{Fe}{2} 4924, 5018, 5169 \AA, with some asymmetry on the blue side that is
consistent with wind.  

Overall, a good match is apparent between the optical \fudesig\ spectra and FU Ori template sources.  

%\begin{figure}[!t]
%\includegraphics[width=0.95\linewidth,trim={0.85cm 0 0 0},clip]{ZTF19abymxrr_spexLM.jpg}
%\caption{
%IRTF/SpeX $LM$ spectrum of \fudesig -- {\bf this is here for discussion purposes, and will be removed since show this spectral range in the model fit figure, below}.
%MC: What is the dip from 4.0 to 4.6 microns?  This FUor is seen through little extinction, as there is no water ice absorption band at 3 microns.  Some FUors that have a strong water ice absorption show a drop off starting around 4 microns, but FUors without water ice absorption just have flat spectra right until the atmosphere becomes opaque at 4.2 um.  Also, a nice JWST spectrum of a high extinction star doesn't show anything like a broad absorption feature in this area.  I don't know if the feature in question is real or uncorrected telluric.  Best not to read much into it until we could get a much higher S/N spectrum of this and other FUors.  
%}
%\label{fig:spexLM}
%\end{figure}

\subsubsection{Infrared}

The near-infrared spectrum of \fudesig\ (Figure~\ref{fig:irspec}) shows atomic lines such as 
the \ion{Fe}{1}, \ion{Mg}{1}, \ion{Si}{1}, and \ion{K}{1} line forest,
several \ion{Si}{1} and \ion{Al}{1} lines, as well as the relatively hot \ion{Ca}{2} lines,
all in the $J$-band. 
In $H$-band, the gravity-sensitive \ion{Si}{1} line and several lines of \ion{Mg}{1}
are present, though less prominently so than in the V960 Mon comparison spectrum.
In the $K$-band, similarly, the diagnostic atomic lines of \ion{Na}{1}, \ion{Ca}{1}, and \ion{Mg}{1} 
are weak, but present.
Clear molecular band absorption is seen throughout the near-infrared in \fudesig, 
e.g. VO, TiO, H$_2$O, and CO, as well as a hint of the surface-gravity sensitive CN band. 
The VO and H$_2$O bands seem stronger than in V960 Mon, whereas CO seems weaker.

Overall, a good match is apparent between the near-infrared \fudesig\ spectra and FU Ori template sources.  

In the mid-infrared spectral range 
%(see the full optical to mid-infrared outburst spectrum in Figure~\ref{fig:modsed}), 
the continuum in the $L$ through $M$ bands is fairly flat.  
There is no evidence of water ice absorption in the 2.8-3.6 $\mu$m region, 
which suggests that the object is seen through little extinction \citep{connelley2018}.  
This is consistent with the finding from the stellar photosphere fitting 
of the pre-outburst SED (above) and the FU Ori accretion disk modelling 
presented below, both of which result in relatively low values 
of estimated $A_V$, in the 1.5-2.0 mag range.

\subsection{Spectral Analysis}

\begin{figure*}[!t]
\centering
\begin{subfigure}[b]{0.48\linewidth}
 \centering
 \includegraphics[width=\linewidth,angle=-90]{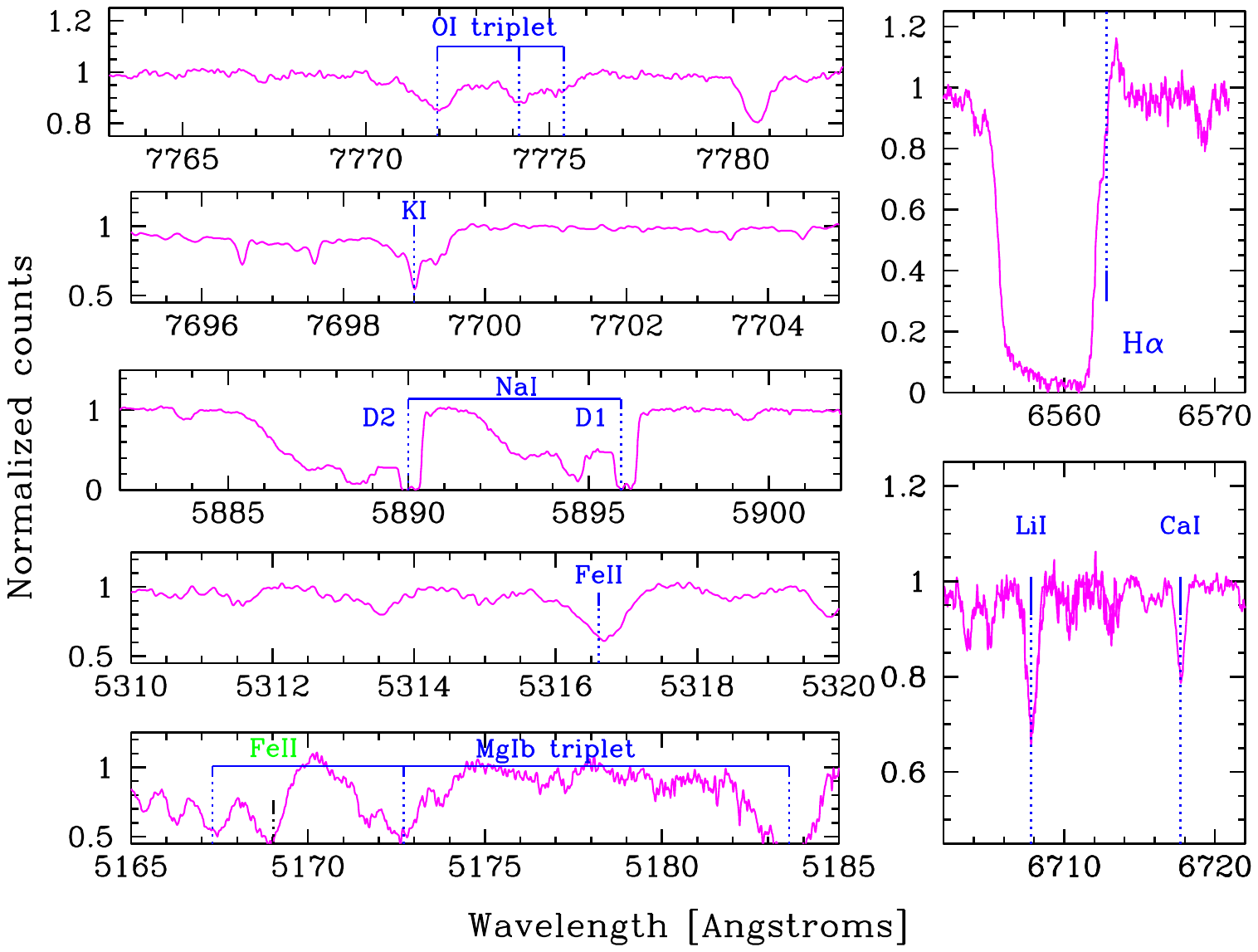}
\end{subfigure}
\hfill
\begin{subfigure}[b]{0.48\linewidth}
 \centering
%\vspace{5cm}
\includegraphics[width=\linewidth,trim={0 0 5cm 5cm},clip]{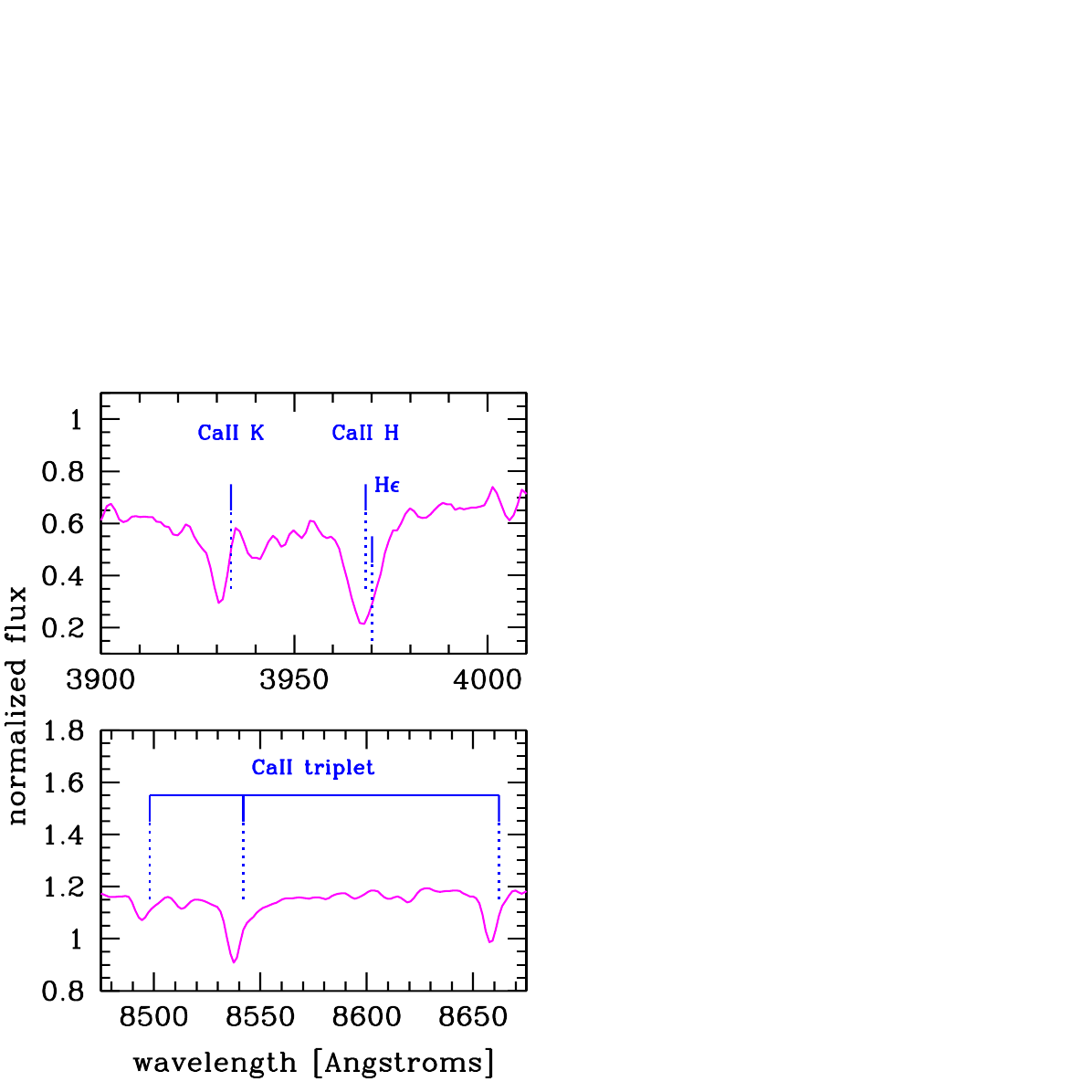}
\vspace{-8cm} 
\hspace{-5cm} 
\end{subfigure}
\caption{
Segments of the Keck/KPF optical spectrum of \fudesig, 
emphasizing notable lines. Strong wind signatures are seen in
H$\alpha$ 
%{\bf add H$\beta$}
which is heavily absorbed (though shows a narrow P Cygni-type emission component), 
and \ion{Na}{1} D doublet, which is saturated. Weaker but significant winds
are also seen in the blueshifted \ion{Mg}{1}b doublet and \ion{Fe}{2} 5316 \AA\ lines. 
The \ion{O}{1} 7773 \AA\ triplet and \ion{K}{1} 7669,7699 \AA\ doublet 
are broad, but not obviously blueshifted.
Also highlighted is the \ion{Li}{1} 6707 \AA\ signature of stellar youth.
Right panels show the \ion{Ca}{2} H\&K doublet and ``infrared" triplet,
both from the lower resolution Palomar/DBSP spectrum.
}
\label{fig:profs}
\end{figure*}

\fudesig\ clearly exhibits a mixed spectrum, having both cool molecular and hotter atomic contributors
to the opacity.  Additionally, we highlight the low surface gravity implied by several optical
(\ion{Ba}{2}, a few \ion{Ti}{1} lines, and the ratio of \ion{Fe}{2} 5316 \AA\ to \ion{Fe}{1} 5328 \AA) lines, as well as near-infrared (\ion{Sr}{2} and CN in the $Y$-band, \ion{Si}{1} line in the $H$-band) signatures. 
\fudesig\ also exhibits the \ion{Li}{1} 6707 \AA\ line, at greater strength 
than the nearby \ion{Ca}{1} 6717 \AA\ line 
(Figure~\ref{fig:profs}), which is an indicator of stellar youth and 
ubiquitously present in FU Ori stars. 

We conclude that the photosphere of \fudesig\ is dominated by absorption lines that appear to be
produced in an accretion disk.   We model this absorption line spectrum in \S~\ref{sec:model}. 

The line widths are narrow, corresponding to broadening of only $v \sin i\approx 15$ km/s, based on $\chi^2$ fitting results for the HIRES spectrum. 
This is not typical of FU Ori type sources which generally have signficant line broadening,
though it is very similar to the spectral appearance of V1515 Cyg.
Narrow lines can be explained in the accretion disk scenario by a near face-on orientation (see \S~\ref{sec:model}).

Such prominent narrow lines in the optical spectrum preclude the unambiguous identification
of diffuse interstellar band (DIB) absorption, which would be a useful diagnostic 
of line-of-sight extinction \citep[e.g.][]{carvalho2022}.  
The strong DIBs features at 5780 and 6614 \AA\ are not clearly distinguishable in \fudesig\ 
amidst other narrow absorption features near these wavelengths, arising in the disk.

Comparing our high dispersion spectra with spectral standards and with model atmospheres 
implies a heliocentric radial velocity $v_{\rm{hel}} = +28.3$ km/s for \fudesig.  
The corresponding $v_{\rm{LSR}}$ is $+20.0$ km/s.
Such a source velocity is consistent with radial motions in this part of the outer Galaxy, 
but unfortunately does not help constrain the source distance. This is due to ambiguities 
in the kinematic models near the anti-center Galactic longitude.

\subsection{Wind/Outflow Lines}

\fudesig\ also exhibits the spectral signatures of outflowing material.
Strong wind and outflow are evidenced by several true P Cygni type lines and 
additional asymmetrically blueshifted, absorption features. 
Figure~\ref{fig:profs} illustrates several such lines in the optical spectrum.

Only H$\alpha$ has a true P Cygni type profile, though with a very modest, narrow emission component. 
While some true P Cygni signatures can be exhibited in FU Ori objects,
blueshifted absorption is the more common wind signature, 
especially in the higher Balmer series lines. 
Wind evidence is also seen in the certain 
metal lines, including the \ion{Mg}{1}b triplet, the \ion{Na}{1} D doublet, 
the \ion{K}{1} 7669,7699 \AA\ doublet, and the \ion{O}{1} 7773 \AA\ triplet.
 
There is no evidence for \ion{Si}{2} 6347,6371 \AA\ (8.1 eV) 
which have presented in some YSO outbursts 
(see e.g. \citealt{hillenbrand2019} on PTF 14jg and \citealt{carvalho2023b} on V960 Mon) 
and are interpreted as hot wind lines.  The \ion{O}{1} 8446 \AA\ (9.5 eV) line 
is present in absorption, but is not strong.  Relatively weak 
higher excitation lines suggest a moderate 
to cool disk and wind relative to other FU Ori stars.

The infrared spectrum of \fudesig\ exhibits a strong blue-asymmetric profile 
in the \ion{He}{1} 10830 line (Figure~\ref{fig:irspec}) 
with line FWHM of about 245 km/s absorption extending out to -400 km/s. 
Pa$\beta$, Pa$\gamma$, and Pa$\delta$ also show wind absorption, 
with line FWHM values around 135 km/s.  These \ion{H}{1} and \ion{He}{1} 
wind lines are much broader than the narrow disk lines, which are unresolved
in our infrared spectra. There is no obvious signature of Br$\gamma$.

\subsection{Spectral Continuum}

To prepare the data for the SED modelling effort described below, 
we note that the Palomar/DBSP, APO/TSpec, and IRTF/SpeX data
are all flux calibrated, but may still have some unaccounted-for slit losses,
requiring absolute offsets in order to rectify them.
From the APO/TSpec data, a photometric brightness  $K_s=10.25$ mag is derived, 
(along with $H= 10.84$ and $J=11.75$ mag)  % c.f. Gattini outburst is 11.3 suggesting still contamination from the bright object to NE
which agrees well with the $K_\mathrm{MKO}=10.31$ photometric measurement 
separately obtained using the IRTF guider, and to which the IRTF/SpeX spectra were normalized.
The optical spectrum from Palomar/DBSP needed to be scaled by about 25\% 
in order to match the near-infrared spectra, likely due to seeing-induced slit loss effects.

\section{Accretion Disk Model}
\label{sec:model}

Above we demonstrated that \fudesig\ displays the spectral signatures of an FU Ori star, 
namely disk-like photospheric absorption between 0.4 and 2.4 $\mu$m, as well as strong wind.
In this section we show that its SED is also well-fit by a standard pure-accretion disk model \citep{SS73}. 
We then optimize the model to derive physical parameters for both the central star and the disk.

The data included in the modelling effort are: the DBSP optical spectrophotometry, 
the SpeX/SXD near-infrared spectophotometry, and the SpeX/LXD mid-infrared spectrophotometry.

\subsection{Fitting Process}

\begin{figure}
\includegraphics[width=\columnwidth]{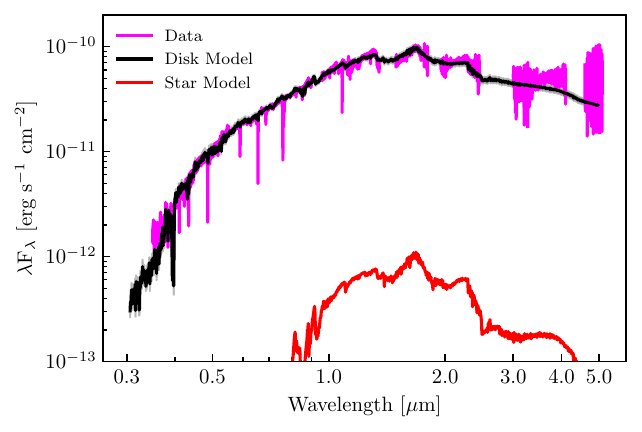}
\includegraphics[width=\columnwidth]{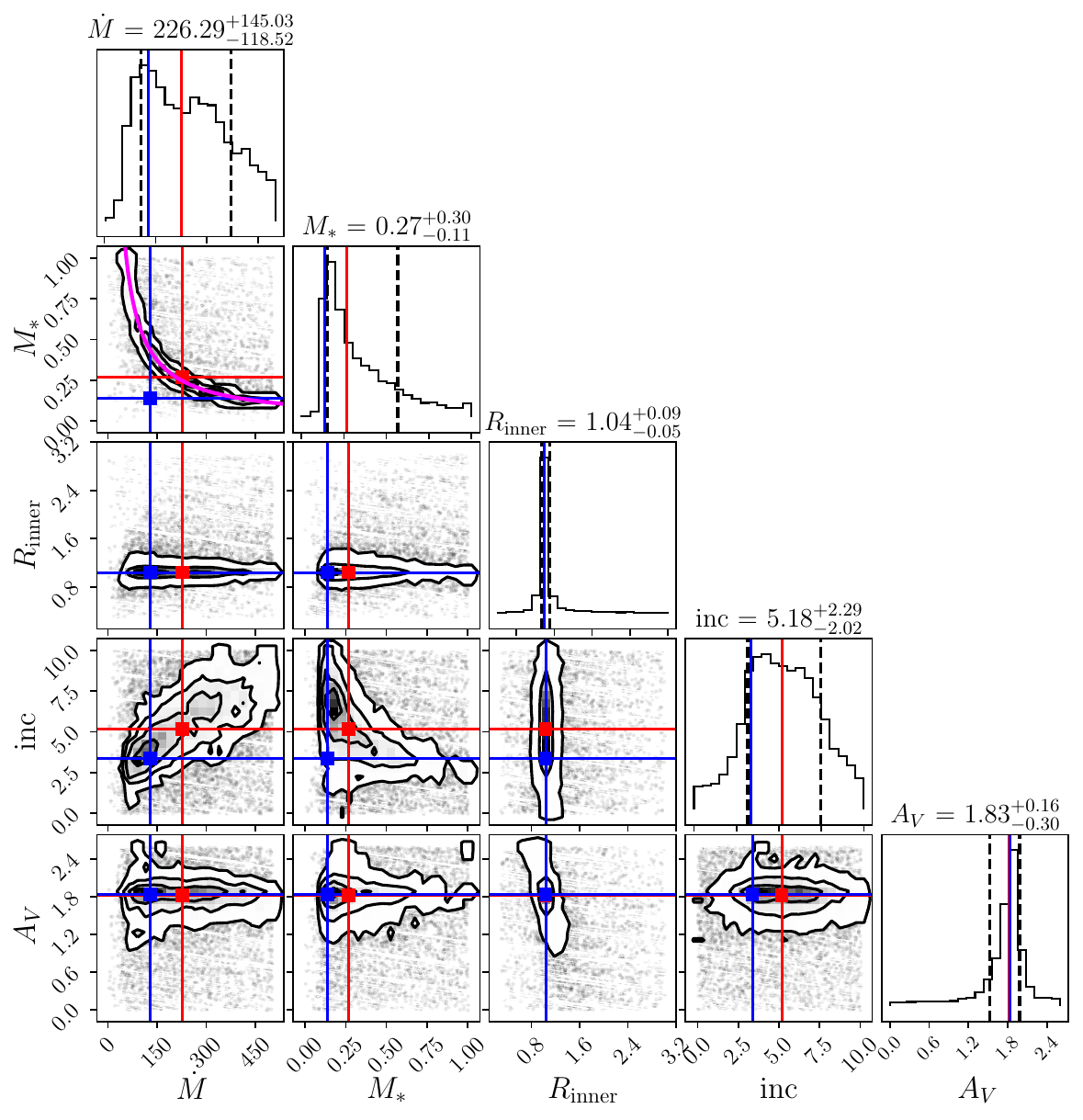}
\caption{Top: Accretion disk model (black) fit by solving for the five parameters 
$\dot{M_\mathrm{acc}}$ (in units of $10^{-8} \ M_\odot$ yr$^{-1}$), $M_{*}$, $R_\mathrm{inner}$, $i$, and $A_V$, 
compared to the full optical through 
mid-infrared spectrum (magenta) assembled from Palomar/DBSP  
and IRTF/SpeX for the outburst source \fudesig.
The estimated progenitor star is shown in red.
Bottom: Corner plot  illustrating the mutual constraints on the various 
physical parameters in the disk model. 
Red and blue lines mark median and modal values. 
The magenta curve in the second-from-top box in the left column
marks the relation $M_* \dot{M} = 5.6 \times 10^{-7} \ M_\odot^2 \ \mathrm{yr}^{-1}$. 
}
\label{fig:modsed}
\end{figure}

To fit an accretion disk model to the outburst spectophotometry of \fudesig,
we followed a procedure similar to that in \cite{hillenbrand2023}
which builds on the infrastructure described in \cite{carvalho2023a} and \cite{carvalho2024a}. The fitting is designed to constrain the variables:
disk accretion rate $\dot{M}$, 
stellar mass $M_*$, stellar radius $R_*$, 
disk inclination $i$, 
and source extinction $A_V$. 

The accretion disk model we adopt is described by the viscously heated thin disk temperature profile \citep{SS73}
\begin{equation} \label{eq:TProf}
    T^4_\mathrm{eff}(r) = \frac{3 G M_* \dot{M}}{8 \pi \sigma r^3} \left( 1 - \sqrt{\frac{R_\mathrm{inner}}{r}}  \right)    ,
\end{equation}
where for $r < \frac{49}{36} \ R_\mathrm{inner}$, we assume the profile becomes isothermal so $T_\mathrm{eff}(r < \frac{49}{36} \ R_\mathrm{inner}) = T_\mathrm{max}$ and $T_\mathrm{max}=T_\mathrm{eff}(49/36 \ R_\mathrm{inner})$ \citep{kenyon1988}. This gives an accretion luminosity of $L_\mathrm{acc} = \frac{1}{2} \frac{GM_* \dot{M}}{R_\mathrm{inner}}$.

We also assume the gas is in Keplerian orbit around the star and that the velocity we observe is projected along our line of sight according to the disk inclination $i$, so that the radial dependence of the velocity is $v(r) \sin i = \sqrt{GM_*/r} \sin i$. The maximum rotational velocity in the disk would then be $v_\mathrm{max} = v(R_\mathrm{inner}) \sin i$. We apply the Keplerian rotational broadening to each annulus in the model via direct integration, as described in \citep{carvalho2023a}.

We fit the SED using a maximum-likelihood MCMC technique with the nested sampling package $\mathtt{dynesty}$ \citep{speagle2020,higson2019}. Uniform distributions of $M_*$, $\dot{M}$, $R_\mathrm{inner}$, $i$, and $A_V$ were sampled. 
For $v_\mathrm{max}$, based on the observed line broadening that corresponds to $v \sin i\approx 15$ km/s in the high-dispersion spectra,  
we imposed a constraint on the log-likelihood of $v_\mathrm{max} =20 \pm 5$ km s$^{-1}$.
Following the recommendation in \cite{carvalho2024b}, based on the relatively blue color $J - Ks < 1.3$ of the \fudesig\ system, and its likely low $L_\mathrm{acc}$ (justified below), we also adopt $R_\mathrm{outer} = 50 \ R_\odot$ for the radius of the outer disk.
Finally, since a precise distance to the source is lacking (\S\ref{sec:pre}), we performed fits for four possible distances: 1, 1.5, 2, and 2.5 kpc
in part to study empirically the relationships of the output parameters to distance.

%To constrain the most important disk parameters we chose to adopt fixed values for $d$, $A_V$, and $i$.  
%Adopted priors include a Kroupa IMF for central star mass, and 
%a gaussian distribution for the $v_\mathrm{max} \times \sin i$ value
%of the innermost annulus in the keplerian disk, 
%centered on 15 km s$^{-1}$ and with a standard deviation of 10 km s$^{-1}$. 

\begin{figure}
\includegraphics[width=0.49\linewidth]{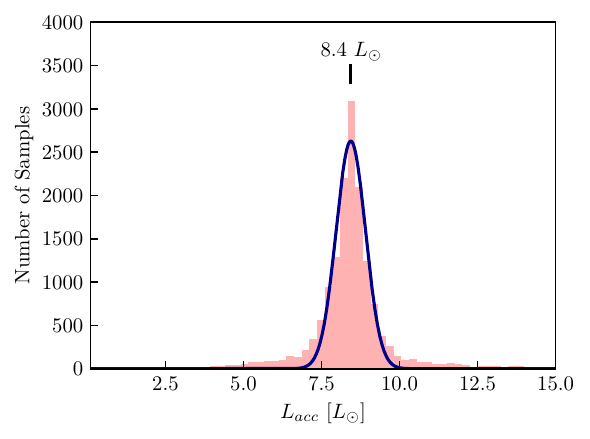}
\includegraphics[width=0.49\linewidth]{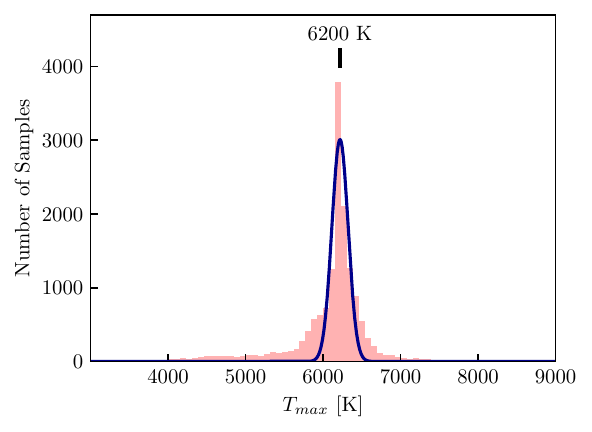}
\caption{Accretion luminosity, $L_\mathrm{acc}$, and maximum disk temperature, $T_\mathrm{max}$, posterior distributions (light red) 
computed from the individual converged MCMC values for the disk model fitting to the observed SED. 
Gaussian fits (dark blue) are used to determine the mean (vertical hashes) and standard deviation for both parameters. 
}
\label{fig:modsedLT}
\end{figure}

\begin{figure}[!b]
\includegraphics[width=0.98\linewidth]{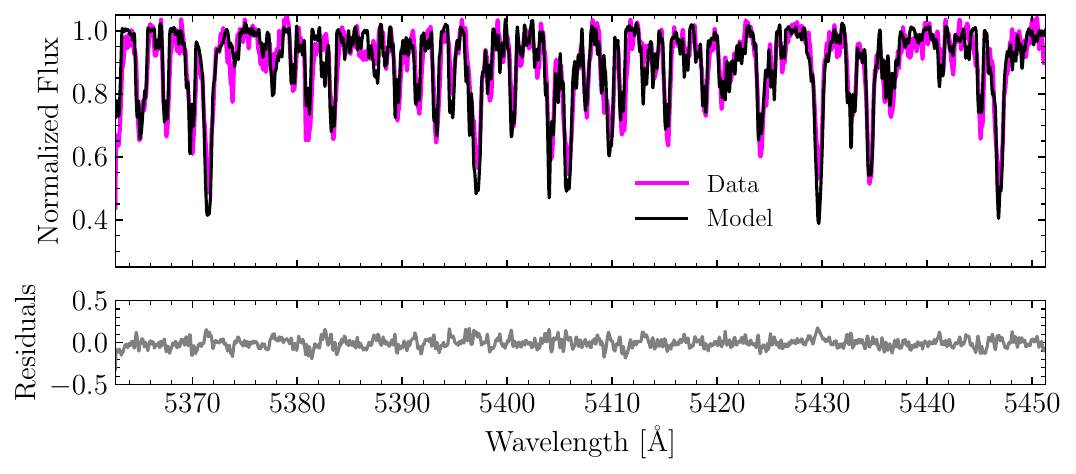}
\includegraphics[width=0.98\linewidth]{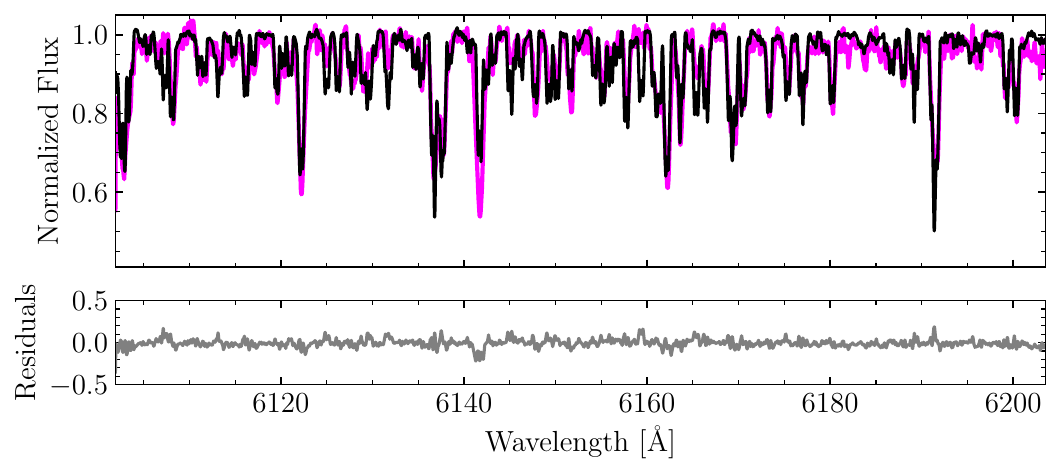}
\includegraphics[width=0.98\linewidth]{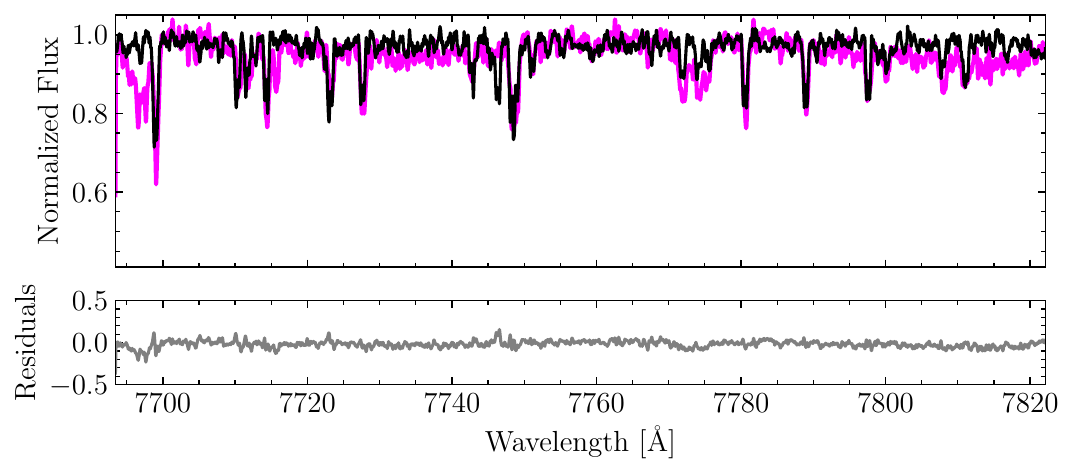}
\caption{Disk model (black) compared to the observed HIRES spectrum (magenta),
along with residuals, shown in three different regions of the optical spectral range.
A superb match is seen between the model and the high dispersion data, 
including line broadening and relative line strengths.
}
\label{fig:hires_diskmod}
\end{figure}

% \pla{In general, the gas disk model is appropriate over the optical and near-infrared wavelength range,
%where accretion dominates the emission. 
%Beginning in the mid-infrared, dust contributions from an inactive or passive disk 
%become important, while in the far-infrared dust emission from an envelope dominates;
%these cooler contributions to the SED are not part of our modelling effort.}

\subsection{Fitting Results}

Results of the above procedures are presented in Figure ~\ref{fig:modsed}.
Then in Figure ~\ref{fig:modsedLT}, we demonstrate how the constraints placed by the model fitting  propagate to posterior distribution for the accretion luminosity $L_\mathrm{acc}$ and maximum temperature of the accretion disk $T_\mathrm{max}$. 

In all of the fits at different distances, we found $i = 5 \pm 3$ deg and $A_V = 1.8\pm 0.2$ mag. The other three parameters, $M_*$, $\dot{M}$, and $R_\mathrm{inner}$, are all functions of the assumed distance.
To better understand the relationship between distance and the best-fit $M_*$, $\dot{M}$, and $R_\mathrm{inner}$, we computed $T_\mathrm{max}$ and $L_\mathrm{acc}$ for each sample of $M_*$, $\dot{M}$, and $R_\mathrm{inner}$ in the converged MCMC. This procedure effectively gives posterior distributions for $T_\mathrm{max}$ and $L_\mathrm{acc}$, even though we did not directly sample in that parameter space. 

We find that $T_\mathrm{max} = 6200\pm 100$ K, regardless of distance. This is also in good agreement with our spectroscopically determined optical spectral type of late F or early G.  %$6000\pm 500$ K. 
Regarding $L_\mathrm{acc}$, in the modelling, the output distributions follow $L_\mathrm{acc} \propto d^2$ exactly, with $L_\mathrm{acc} = 8.4 \pm 0.5 \ L_\odot$ at 1.5 kpc.  The tightly constrained $T_\mathrm{max}$ and distance-dependent $L_\mathrm{acc}$ immediately give us the relationships between distance and the remaining physical parameters. Since $T_\mathrm{max}^4 \propto L_\mathrm{acc}/R_\mathrm{inner}^2$ and $L_\mathrm{acc} \propto d^2$, then $R_\mathrm{inner} \propto d$, which we see in the best-fit values (within parameter uncertainties). Similarly, since $T_\mathrm{max}^4 \propto M_* \dot{M}/R_\mathrm{inner}^3$, the product $M_* \dot{M} \propto d^3$. And since $v_\mathrm{max}^2 \propto M_*/R_\mathrm{inner}$, $M_* \propto d$ so $\dot{M} \propto d^2$.

Adopting a distance $d=1.5$ kpc, the best-fit system parameters become: 
$M_* = 0.17 \pm 0.07 \ (d/ 1.5)$ $M_\odot$; 
$\dot{M} = 10^{-5.48 \pm 0.04} \ (d/ 1.5)^2 \ M_\odot$ yr$^{-1}$; 
$R_\mathrm{inner} = 1.04 \pm 0.1 \ (d/1.5) \ R_\odot$. The $M_*$ posterior distribution is skewed, so we adopt the modal value as the best-fit and approximate the uncertainty based on the FWHM of the distribution of samples around 0.17 $M_\odot$. For the other parameters we adopt the median value as the best-fit and use the 16th and 84th percentile sample values as the $\pm 1\sigma$ uncertainties. 
If the inner disk radius $R_\mathrm{inner} \sim R_*$, the stellar radius,
then the implied $M_*$ and $R_*$ values are approximately consistent 
with the \citet{Baraffe_isochrones_2015A&A} 2-3 Myr isochrones.
Next, we compare this model to the observational details of \fudesig.

\subsection{Comparison of Fitting Results to Observations}

The best-fit accretion disk model that is described above provides a good match 
to the available spectrophotometry of \fudesig,  as shown in Figure~\ref{fig:modsed}. 
We note that the mid-infrared ($LM$) flux is almost entirely reproduced  by the active inner disk,
with the innermost regions of the larger-scale passive disk or thin envelope 
that are evidenced in the SED, contributing very little.
This situation is in contrast to other FU Ori systems, 
such as V960 Mon, HBC 722, and RNO 54, where the flux in the $LM$ region is 
insufficiently reproduced by the inner outbursting disk, and more substantially
attributed to the non-active part of the disk at larger disk radii 
\citep{carvalho2023a,carvalho2024a,hillenbrand2023}. 

Figure~\ref{fig:hires_diskmod} shows the same accretion disk model relative to the
observed HIRES spectrum for three different spectral segments.
Overall, the disk model is an excellent fit to the data, including for
the abundant lower-exitation lines, as well as for higher-excition lines like
\ion{O}{1} 7773, 8446 \AA, \ion{Ca}{2} 8912, 8927 \AA, and several
\ion{C}{1} lines between 9061 and 9111 \AA.
The presence of these hotter lines reinforces the concept of the disk-like
atmosphere, with a sufficiently high $T_\mathrm{max}$ at the inner edge.
The main discrepancies are in \ion{Li}{1} which is not well-represented
by standard atmosphere models, and in the TiO feature at 8860 \AA\ as 
discussed in previous literature \citep{herbig2003,carvalho2023b}.
%as discrepant between FU Ori disks and model atmospheres.

There are also wind-affected lines, specifically  
\ion{Na}{1} D, H$\alpha$, \ion{K}{1}, and the \ion{Ca}{2} triplet where the
disk model is not a good fit. For these and other such wind-affected lines, 
the model can be used to isolate the wind profile.
%{\bf [also a 9160A broad feature, which seems like an artifact]}

Notably several hot lines in the 5-10 eV range that are seen in 
the optical and near-infared of some FU Ori objects 
with high values of $T_\mathrm{max}$, are absent or only very weakly present in \fudesig.  
These include various \ion{Si}{1}, \ion{C}{1}, \ion{Mg}{2} and \ion{Ca}{2}
absorptions that are exhibited by sources such as RNO 54 ($T_\mathrm{max} \geq 7000$ K; \citealt{hillenbrand2023}).
With a lower $T_\mathrm{max} \approx 6200$ K, \fudesig\ may not be expected to show these higher 
temperature species.

Overall, we find the pure-accretion disk model a good match from the ultraviolet/blue
to mid-infrared spectrophotometry of the outburst source \fudesig.  The same disk model
also reproduces well the optical high dispersion spectroscopy.

\begin{table*}
\centering
\caption{Properties of FUOr-Aur 0544+3330}
\label{tab:properties}
\begin{tabular}{|crc|}
\hline   
\hline   
                               Variable &                               Value &         Unit \\
\hline   
\multicolumn{3}{|c|}{\bf [Measured and Modelled Parameters]} \\ 
                Right Ascension (R.A.) &                                         05:44:52.25 & J2000.\\
                Declination   (Dec.) &                                           +33:30:09.6 & J2000.\\
               R.A. Proper Motion ($\mu_{R.A.}cos(Dec.)$) &  -2.37 & mas/yr \\
               Dec. Proper Motion ($\mu_{Dec.}$) & +0.17  & mas/yr \\
                Radial Velocity ($v_{hel}$) &                                         +28.3 & km/s \\
\hline   
%               Outburst Amplitude ($\Delta g$)&                                   -5......&  \\
                Outburst Amplitude ($\Delta r$)&                                   -5.1 & mag \\
                Outburst Amplitude ($\Delta J$)&                                    -3.9 & mag \\
                Outburst Amplitude ($\Delta H$)&                                    -3.9 & mag \\
                Outburst Amplitude ($\Delta K$)&                                 -3.5 & mag \\
                Outburst Amplitude ($\Delta W1$)&                                  -2.7 & mag  \\
                Outburst Amplitude ($\Delta W2$)&                                  -2.5 & mag  \\
\hline   
                     Extinction ($A_V$) &                                            1.8 & mag\\
         Disk           Inclination (i) &                                             6 & deg\\
         Disk Projected Broadening    (v~$\sin(i)$) &                                            15 & km/s \\
         Disk  Maximum Temperature ($T_{max}$)& 6218 & K  \\
Disk Accretion Rate (log $\dot{M}$)& -5.48 & dex $M_\odot$ yr$^{-1}$ \\
Disk Accretion Luminosity ($L_{acc}$)                        & 8.4 & L$_\odot$ \\
         Disk  Inner Radius ($R_{inner}$)&                                        1.0 & R$_\odot$  \\
                Stellar Mass ($M_*$) & 0.17 & $M_\odot$ \\
               Mass Energy Flux ($M_* \dot{M}$) & $5.6 \times 10^{-7}$ & $M_\odot^2 \ \mathrm{yr}^{-1}$ \\
\hline   
\multicolumn{3}{|c|}{\bf [Estimated and Adopted Parameters]} \\ 
                         Distance (d) &                                          1500 & pc   \\
    Progenitor   Temperature     ($T_{eff}$) &                                   3100 & K       \\
    Progenitor   Luminosity      ($L_*$) &                                   0.55 & L$_\odot$\\
    Progenitor   Surface Gravity (log g) &                                     4.0 & dex cm/s$^2$ \\
          Disk Outer Radius ($R_{outer}$)&                                        50 & R$_\odot$\\
\hline   
\hline   
\end{tabular}
\end{table*}

\section{Discussion}
\label{sec:disc}
\subsection{A new naming system for FU Ori objects}
We begin our discussion by proposing a novel naming system for newly discovered or newly confirmed FU Ori objects.
The scheme uses ``FUOr-" to indicate the FU Ori class of the source, followed by the traditional
constellation name, then a shortened form of the source coordinates in
hours of right ascension and positive/negative degrees of declination.
The new FU Ori object that we present here is thus designated \fudesig, 
whereas the prototype FU Ori itself would be called FUOr-Ori 0545+0904.
%{\bf[alternates: FO-Aur?  FOO-Aur?  FUOO-Aur?]}

We further suggest the requirement that 
both an outburst lightcurve and a multi-temperature spectrum need to be demonstrated 
in order for an object to be assigned an FUOr-[constellation][RA][Dec] name.
We also re-iterate the emprical criteria of \cite{connelley2018} for FU Ori status, with which we agree:
``deep CO band absorption, weak metal absorption lines, 
pronounced water vapor at the short and long sides of the H-band window 
giving rise to a characteristic triangular H-band continuum, a strong break at 1.32 $\mu$m due to water vapor,
strong He I absorption (frequently blueshifted) at 1.083 $\mu$m, and few if any emission lines."
These authors also mention association with a star forming region, and the presence of \ion{Li}{1} absorption,
for optically visible sources, as desirable characteristics.

\subsection{\fudesig\ as an FU Ori Object}

The hypothesis of an accretion outburst from the young star progenitor \prog,
that was detected as photometric alert \obj, and has become the outburst source \fudesig,
is supported by several lines of evidence.  This includes the Class I-like YSO nature 
of the progenitor SED,
the detected outburst at multiple optical and infrared wavebands, the spectroscopic
similarity to other outbursting FU Ori objects including strong wind signatures, 
and the excellent fit of the outburst SED to a pure-accretion disk model.   

\fudesig\ joins a growing list (Contreras Pena 2025, in preparation) %\cite{ccp2025} 
of mature FU Ori objects and recently detected outbursts.  Sources displaying characteristics
similar to those of \fudesig\ are commonly accepted to be in a state of prolonged,
enhanced or ``high-state" accretion.  In such sources, the protoplanetary disk 
surrounding the young star has increased the disk-to-star mass transfer,
from $\sim 10^{-11}$ to $10^{-8}\ M_\odot$ yr$^{-1}$ in the low state of normal T Tauri accretion,
to levels as high as $10^{-5}$ to $10^{-4}\  M_\odot$ yr$^{-1}$ in the high state or outburst. 
Many authors \citep[e.g.][]{kenyon1988,welty1992,rodriguez2022,carvalho2023a,liu2022}, 
have demonstrated that the high-dispersion spectra of FU Ori stars
can be adequately reproduced by pure-accretion disk models. We have done so here, with the 
model's temperature gradient and velocity gradient with radius combining to produce
the observed absorption spectrum from blue optical through the near-infrared for \fudesig.

\fudesig\ is a particularly valuable member of the FU Ori class because
its photometric rise was sampled in at least seven wavebands (Figure~\ref{fig:lc}).
Furthermore, \fudesig\ enjoys decent sampling of its lightcurve in the pre-outburst 
stage, which enables accurate sampling of the rise shape, as well as measurement of the full 
rise amplitudes and timescales -- from blue optical to mid-infrared.

%Fewer than {\bf XXX of the $\sim$30 well-accepted} {\it bona fide} FU Ori objects \citep{ccp2025}  have been observed -- in any waveband -- as they underwent their dramatic brightness increases.  The rest have been classified as likely FU Ori objects only after the fact,  based on late-time spectroscopy. 
%\pla{
%For example,
%V582 Aur (Munari et al. 2009; Samus 2009) was relatively recently recognized 
%as a member of the class, in 2009, but likely outburst in the mid- 1980s (Semkov et al. 2013).
%RNO 54 \citep{hillenbrand2024} is similarly, relatively recently recognized.

%Given the recent discovery rate of approximately one new FU Ori outburst every 1-2 years,
%and the lower rates in the past (one discovery only every $\sim$10 years),
%there are likely to be a few tens of unrecognized ``FUOr-like" stars
%in which no outburst was detected/noticed, but the object currently does
%exhibit the spectral signatures of being in an outburst state.
%}

Compared to other FU Ori stars that are bright enough for optical high dispersion spectroscopy, 
\fudesig\ seems to have narrow absorption lines. 
This is due to its very low inclination, i.e. near to face-on orientation. 
The closest analog among well-studied FU Ori sources is V1515 Cyg;
V2775 Ori also exhibits similarly narrow lines in the near-infrared (Carvalho 2025, in preparation).

\fudesig\ is one of a small number of low luminosity FU Ori objects. 
Other sources having $L < 20 L_\odot$, presumed entirely due to accretion, 
include L222 78 \citep{guo2024}, PGIRN 20dci \citep{hillenbrand2021}, Gaia 17bpi \citep{hillenbrand2018}, 
and V2495 Cyg \citep{movsessian2006,connelley2018}.  While the luminosity may be low, 
the accretion rate can still be high or moderate, as is the case for \fudesig\ 
with $\dot{M}_{acc} = 3.3\times10^{-6}$,  if the central star mass $M_*$ or $M_*/R_*$ is low. As $M_*$ and $R_*$ are related by $R_* \sim M_*^{0.5}$ \citep{Baraffe_isochrones_2015A&A, nayakshin2024}, lower mass stars should naturally experience somewhat lower $L_\mathrm{acc}$ and $\dot{M}$ outbursts.
Thus high luminosity is not a requirement for FU Ori status, though it has been a common characteristic 
of the objects historically identified as FU Ori outbursts.

\fudesig\ is also one of a small number of identified FU Ori objects that are 
located either in small dark clouds or in remote areas on the peripheries
of known star forming regions, but not near their centers where
star formation is perceived as being most active.
Other examples of relatively isolated FU Ori objects are: Gaia 17bpi \citep{hillenbrand2018}, RNO 54 \citep{hillenbrand2023}, and V582 Aur \citep{kun2017}. 
The last of these is only a few degrees away from \fudesig. % 05 25 51.974 +34 52 30.10  

\subsection{The binarity of \fudesig}
It remains unclear whether the dominant triggering mechanisms for FU Ori outbursts
are driven internally, by unstable disk physics, or externally through dynamical interactions 
\citep[e.g. see recent review material in][]{nayakshin2024}.
In the case of \fudesig, high spatial resolution imaging is still needed in order to search for
the companion that we have hypothesized as dominating the optical light from the progenitor source
\prog, as illustrated in the SED (Figure~\ref{fig:obssed}).   
If such a companion could be found, the source would be similar to the prototype 
FUOr-Ori 0545+0904 (FU Ori itself) in that the outbursting source is 
the less massive component of a moderate-separation, few hundred AU binary.

We infer the presence of a binary component based on the difficulty of fitting the pre-outburst spectrum with a single star spectrum and the tension between the best-fit pre-outburst system temperature and the best-fit properties of the central object during outburst. The pre-outburst optical SED is relatively flat and thus indicates that any molecular absorption expected from cooler photospheres should be minimal. Even assuming strong accretion veiling the absorption lines, this places a lower limit of $\sim 3500$ on the pre-outburst central object temperature. Such a low temperature (and therefore low mass) object would be severely underluminous for the adopted 1.5 kpc distance. 

Adopting a closer distance does not remedy this because for a 2 Myr system the intrinsic $L_* \propto M_*^{3/2}$ \citep{Baraffe_isochrones_2015A&A}, which adopting the outburst best-fit mass dependence on distance gives $L_* \propto d^{3/2}$, while the observed preoutburst luminosity $L_\mathrm{pre} \propto d^2$. Varying the distance will not bring the intrinsic luminosity of the best-fit central star and the measured luminosity of the pre-outburst object into agreement for $d>100$ pc. Another strong constraint on the pre-outburst object is the tight posterior distribution of $R_\mathrm{inner}$. A central star more massive than $0.4 \ M_\odot$, which could help to remedy the pre-outburst luminosity problem, is ruled out for a system younger than 5 Myr \citep{Baraffe_isochrones_2015A&A}. For $M_* < 0.4 \ M_\odot$, $L_*< 0.2 \ L_\odot$, which is much lower than the desired $5 \ L_\odot$ measured preoutburst.

\subsection{Testing the bolometric correction method of luminosity estimation}
Finally, we can test the method promoted by \cite{carvalho2024b}
of using bolometric corrections in order to derive FU Ori accretion disk luminosities.  Above, our detailed SED modelling resulted in $L_\mathrm{acc} \approx 8.4 \ L_\odot$ for \fudesig.
Following the bolometric correction prescriptions and
assuming the same source distance and extinction values as above,
we find accretion luminosities of 7.2, 5.9, and 7.8 $L_\odot$, 
from the $J$-band, $W1$-band, and $W2$-band magnitude measurements at MJD = 60343.966. 
All three values are comparable to the more rigorous assessment.

\section{Summary}

We have presented compelling evidence that \fudesig\ is a newly discovered example of the FU Ori phenomenon
and have summarized our findings on the system in Table~\ref{tab:properties}.

The progenitor to the outburst had a Class I type SED with estimated distance 1500 pc.
We speculate that the outburst source may be the secondary in a young binary system.
A photometric outburst was observed across multiple wavelengths
from the optical to mid-infrared.  Lightcurves rose between late 2019 and early 2021 by 
$\Delta r = -5.1$ mag 
%$\Delta J = -3.9$ mag, $\Delta K = -3.5$, $\Delta W1 = -2.7$ mag, and $\Delta W2 = -2.5$ mag, 
and $\Delta W2 = -2.5$ mag, and they became bluer in both the optical and the infrared. 
We have introduced a new functional form to the field, that fits the early shallow rise phase of the source,
the subsequent steeper rise to peak, and the lightcurve turnover: 
$a~\mathrm{tanh}((t - b) / c) \times e^{-(t - b) / d} + \mathrm{offset}$.  

Follow-up spectra and spectrophotometry between $0.36-5.0 \mu$m  were obtained in fall of 2024. 
These data exhibit the expected multi-temperature spectrum,
with a wavelength-dependent spectral type that changes from earlier at optical wavelengths, to later
in the near-infrared.  At high-dispersion, even over short wavelength ranges, there is 
evidence of a mixed-temperature and low-gravity spectrum.  The source also exhibits
deep and broadly blue-shifted absorption lines in species known to trace disk winds in heavily accreting YSOs.
Finally, the outburst spectrum exhibits the \ion{Li}{1} 6707 \AA\ line that is a signature of stellar youth.

We have applied a pure-accretion disk model to the spectrophometry 
from blue optical to mid-infrared wavelengths, and find the  following physical system parameters:
$\dot{M} = 10^{-5.48\pm0.04}$ $M_\odot$ yr$^{-1}$, $M_* = 0.17\pm0.07 \ M_\odot$, and $R_\mathrm{inner} = 1.04\pm0.10 \ R_\odot$.
The corresponding radiative parameters of the accretion disk are thus
$T_\mathrm{max} = 6218 \pm 100$ K and $L_\mathrm{acc} = 8.4 \pm 0.5 \  L_\odot$.
These same parameters produce an excellent match to the high dispersion optical spectroscopy.
We also constrain disk inclination $i=5.9^{+2.3}_{-2.0}$ deg and source extinction $A_V=1.83^{+0.16}_{-0.30}$ mag. 

\begin{acknowledgments}
We thank our referee for detailed comments that helped improve the manuscript.
We acknowledge the contributions of Jack Lubin, %for his role in scheduling KPF observations, 
BJ Fulton, %as the DRP lead 
and Andrew Howard %as CPS lead
-- all within the CPS collabation -- for enabling us to show a KPF version of Figure~\ref{fig:profs}. The work of DS was carried out at the Jet Propulsion Laboratory, California Institute of Technology, under a contract with NASA.
\end{acknowledgments}

\vspace{5mm}
\facilities{Palomar:(DBSP), Keck:I(HIRES), Keck:I(KPF), APO(TSpec), IRTF(SpeX)}

\bibliography{ms}{}
\bibliographystyle{aasjournal}

\appendix

\begin{figure}[h]
    \centering
    \includegraphics[width=0.48\linewidth]{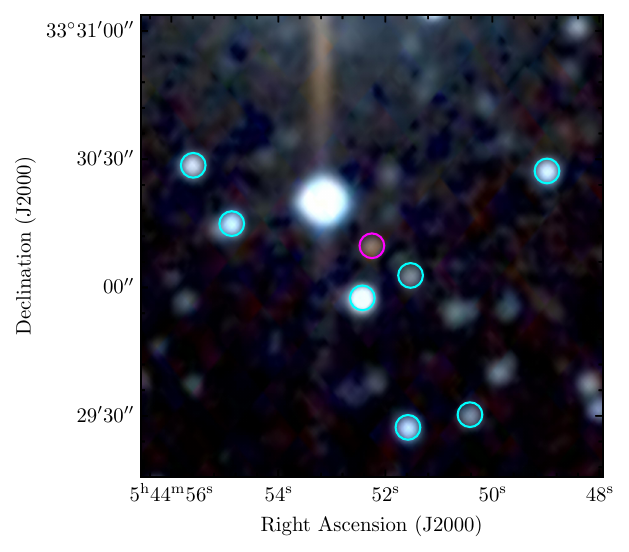}
    \caption{The 2MASS color image showing the neighborhood of \fudesig. The progenitor is marked by the magenta circle and the reference stars used for the differential flux calibration are marked by cyan circles set to the 3$^{\prime\prime}$ 
    diameter aperture used for photometry extraction.
    A bright star off-frame to the north causes a visible diffraction artifact that presumably affected both the 2MASS PSC absence and the 2MASS XSC brightness overestimate for this source.}
    \label{fig:2MASSField}
\end{figure}

\section{2MASS Photometry}
\label{app:2MASS}

The existing cataloged $J,H,K_s$ fluxes for the pre-outburst source appear in the 2MASS Extended Source Catalog \citep[XSC,][]{skrutskie2006}. The source is flagged as being impacted by a diffraction spike from a nearby bright star and is therefore not included in the Point Source Catalog (PSC). However, in the XSC, the radius of the aperture used to extract the flux is 9$^{\prime\prime}$, 
which captures light from neighboring objects. Therefore the reported $J = 14.529 \pm 0.092$, $H = 13.730 \pm 0.098$, $K_s = 13.391 \pm 0.121$ are overestimates of the pre-outburst source brightnesss. Since visual inspection reveals that the source is not impacted by the diffraction spike, we perform our own point-source aperture photometry to compute the NIR SED of \fudesig\ pre-outburst.

We use the $\mathtt{photutils}$ Python package to identify point sources in the image, define apertures for extraction, and to compute the total fluxes within the chosen apertures. The diameter of the aperture used for each band is the estimated $FHWM$ using the geometric factor $\mathtt{seesh}$ given in the 2MASS Atlas image header. We use the recommended formula $FWHM = 3.13 \times \mathtt{seesh} - 0.46$\footnote{\url{https://irsa.ipac.caltech.edu/data/2MASS/docs/supplementary/seeing/seesum.html}}, which gives 2.98$^{\prime\prime}$, 2.89$^{\prime\prime}$, and 2.89$^{\prime\prime}$, for $J$, $H$, and $Ks$, respectively. We adopt the estimated sky background level for each image, since the field is too crowded to perform reliable local background subtraction for each source. 

We calibrate the background-subtracted flux of the \fudesig\ progenitor differentially by comparing it with six neighboring stars that have reported PSC fluxes. The six measurements were averaged to produce a final flux for the progenitor, and we adopt the standard deviation of the measurements as the uncertainty on the flux measurement. Our differential photometry yields: $J = 15.67 \pm 0.03$, $H= 14.73 \pm 0.02$, and $K_s = 13.80 \pm 0.03$.  The field, with both progenitor and reference sources marked, is shown in Figure \ref{fig:2MASSField}.

\section{SED fitting}
\label{app:sedfitter}

We use the Python package $\mathtt{sedfitter}$ \citep{robitaille2007} to fit the observed SED of \fudesig\ to a library of pre-computed spectral energy distribution models for YSOs \citep{robitaille2017}, then filter the results and sort by $\chi^2$ values. 

The specific library of SED models was the $\mathtt{sp-h-i}$ model set, containing 10,000 models of stars with a passive disk and variable inner radius. To conduct the fitting, the 18 photometric measurements ranging from 0.5 to 22 $\mu$m were used, adopting apertures of 3 arcseconds. We note the lack of observational constraints at long wavelengths, and thus on the outer disk.  The source distance was constrained to 1-2 kpc, and the $A_V$ range to 0-40 mag. Selecting data values with a $\chi^2$ no more than 6 above the best $\chi^2$ resulted in 351 remaining models.  

A histogram of the inclination angles of these best-fitting models peaks towards low inclination.
Given the certainty that the source is viewed close to face-on, all models with $i>10$ deg and $\chi^2 > 80$ were eliminated. 
For the 20 remaining models, the $R_\mathrm{outer}$ values were strongly peaked toward low values,
and two with $R_\mathrm{outer}> 1,000$ AU were also eliminated, leaving 18 best-fitting models remaining.

The best-fitting model, having $\chi^2 = 24$, is shown in the figure below.
It is meant to be illustrative, rather than definitive, but appears to have reasonable parameters for a disk surrounding a low-mass YSO.
The $R_\mathrm{inner}=3.2 R_*$ and $R_\mathrm{outer}=90$ AU, with surface density power law index $p=-0.6$ and $\beta=1.3$.
The dust mass is fairly low at $2.5\times10^{-7} M_\odot$, though the SED is inadequate at the long wavelengths  ($>30 \mu$m) where this parameter can be probed.  The inclination is $6$ deg, identical to our accretion disk model fit results.
%Interestingly, the $A_V$ for all 18 models is around 6 mag, somewhat higher than the $A_V \approx 2$ mag found in the outburst. This is plausible in the scenario where the outburst reduces line-of-sight extinction.  However, the higher Av is probably a trade for the high T\* though

\begin{figure}[h]
    \centering
    \includegraphics[width=0.48\linewidth]{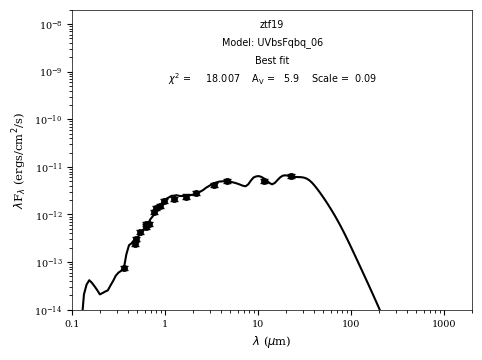}
   \caption{Output format of $\mathtt{sedfitter}$ illustrating the YSO model fit described in the text.}
   %\caption{[apparently we can not access the models, only these default plots of them]}
   \label{fig:sedfitter}
\end{figure}

\end{document}